\documentclass[%
 reprint,
 amsmath,amssymb,
 aps,
]{revtex4-2}

\usepackage{graphicx}
\usepackage{dcolumn}
\usepackage{bm}
\usepackage{hyperref}

\usepackage{upgreek}
\usepackage{makecell} 

\begin{document}

\preprint{APS/123-QED}


\title{Microwave loss characterization using multi-mode superconducting resonators}

\author{Chan U Lei,\normalfont\textsuperscript{1,\,2}}
\author{Suhas Ganjam,\normalfont\textsuperscript{1,\,2}}
\author{Lev Krayzman,\normalfont\textsuperscript{1,\,2}}
\author{Archan Banerjee,\normalfont\textsuperscript{1,\,2}}
\author{Kim Kisslinger,\normalfont\textsuperscript{3}}
\author{Sooyeon Hwang,\normalfont\textsuperscript{3}}
\author{Luigi Frunzio,\normalfont\textsuperscript{1,\,2}}
\author{Robert J. Schoelkopf,\normalfont\textsuperscript{1,\,2}}
\affiliation{\normalfont\textsuperscript{1} Department of Physics and Applied Physics, Yale University, New Haven, Connecticut 06511, USA}
\affiliation{\normalfont\textsuperscript{2} Yale Quantum Institute, Yale University, New Haven, Connecticut 06520, USA}
\affiliation{\normalfont\textsuperscript{3} Center for Functional Nanomaterials, Brookhaven National Laboratory, Upton, NY, 11973, USA}

\date{\today}

\begin{abstract}
Measuring the losses arising from different materials and interfaces is crucial to improving the coherence of superconducting quantum circuits. Although this has been of interest for a long time, current studies can either only provide bounds to those losses, or require several devices for a complete characterization. In this work, we introduce a method to measure the microwave losses of materials and interfaces with a single multi-mode superconducting resonator. We demonstrate a formalism for analyzing the loss sensitivity of multi-mode systems and discuss the design strategies of multi-mode resonators for material loss studies. We present two types of multi-mode superconducting resonators for the study of bulk superconductors: the forky whispering-gallery-mode resonator (FWGMR) and the ellipsoidal cavity. We use these resonators to measure the surface dielectric, conductor, and seam losses of high-purity (5N5) aluminum and aluminum alloy (6061), as well as how they are affected by chemical etching, diamond turning, and thin-film coating. 
We find that chemical etching and diamond turning reduce both the surface dielectric and conductive losses of high-purity aluminum, but provide no appreciable improvement to the seam. Coating the surfaces of diamond-turned aluminum alloys with e-beam evaporated or sputtered aluminum thin-films significantly reduces all three losses under study.
In addition, we study the effect of chemical etching on the surface of high-purity aluminum using transmission electron microscopy (TEM) and find that the chemical etching process creates a thinner and more uniform oxide layer, consistent with the observed improvement in the surface dielectric loss.
\end{abstract}

\maketitle

\section{Introduction}
Improving coherence of superconducting quantum circuits is critical to improving the performance of quantum computers. Many works have shown that the coherence of superconducting quantum circuits is limited by dissipation from their constituent materials \cite{reagor2013reaching, wang2015surface, chu2016suspending, brecht2017micromachined, lei2020high,  calusine2018analysis, woods2019determining, melville2020comparison, kudra2020high, mcrae2020dielectric, heidler2021non, altoe2022localization}. Understanding the mechanism of these losses is crucial to improving coherence. An important step toward this goal is quantifying the microwave losses in the relevant power and temperature regime, and correlating them with the physical properties of the materials. To this end, devices such as superconducting microwave resonators and superconducting qubits are very useful tools because their losses can be measured to very high precision and are highly sensitive to intrinsic material loss. In addition, sensitivities to specific loss channels can be engineered by modifying the geometry of the resonators. Using devices with carefully designed geometries, the microwave losses of the materials can be extracted from the correlations between the energy participations in different loss channels and device relaxation times. This method has been widely used to study various types of microwave loss in superconducting circuits\cite{pappas2011two,reagor2013reaching,minev2013planar,quintana2014characterization, wang2015surface,chu2016suspending,gambetta2016investigating,calusine2018analysis,brecht2017micromachined,woods2019determining,melville2020comparison,romanenko2020three,kudra2020high,mcrae2020dielectric,lei2020high,mcrae2021cryogenic,heidler2021non,checchin2021measurement,altoe2022localization}, such as dielectric losses from substrates and material interfaces\cite{wang2015surface, chu2016suspending, calusine2018analysis, woods2019determining, melville2020comparison, read2022precision, kudra2020high, crowley2023disentangling}, the surface resistance of superconductors \cite{minev2013planar, reagor2013reaching}, losses from the joints between superconductors \cite{Brecht2015, lei2020high}, etc. 

Here, we introduce an alternative approach to quantifying microwave losses of materials using multi-mode superconducting microwave resonators. Instead of using multiple devices with varying geometries to extract losses, we use the multiple modes of a single resonator to extract the microwave losses of all the relevant loss channels at once. Since all the modes are from the same device and measured in the same thermal cycle, this method eliminates the uncertainties from sample-to-sample variations and the possible variations generated by different thermal cycles, e.g., different ambient magnetic fields or thermalization conditions, improving the measurement sensitivity of the system.

We present two types of multi-mode resonators, the forky whispering-gallery-mode resonator (FWGMR) and the ellipsoidal cavity. These devices have a variety of modes that have different sensitivity to loss mechanisms such as surface conductive loss, surface dielectric loss, and seam loss. 
We demonstrate the multi-mode approach to loss characterization by measuring the microwave losses of high-purity (5N5) aluminum and aluminum alloy (6061), which are materials commonly used to make superconducting cavities and enclosures in superconducting quantum devices \cite{ofek2016extending, chou2018deterministic, zhou2021modular, chakram2021seamless}. 
With this technique, we analyze how these intrinsic losses change with different surface treatments such as chemical etching \cite{reagor2013reaching, kudra2020high, chakram2021seamless}, diamond-turning, and aluminum thin-film coating \cite{kuhr2007ultrahigh, benvenuti1999study, roach2012niobium}. In addition, we use transmission electron microscopy (TEM) to identify physical signatures of loss and correlate them with improvements in microwave quality due to chemical etching. 

The outline of this paper is as follows. We begin by introducing the loss model of multi-mode cavity resonators in Sec. \ref{sec:loss model} and describe how to characterize the microwave loss of materials using a multi-mode system. In Sec. \ref{sec:FWGMR} and Sec. \ref{sec:ellipscav}, we present the FWGMR and the ellipsoidal cavity, respectively, and analyze their measurement sensitivity using the method described in Sec. \ref{sec:loss model}. We then use these multi-mode cavity resonators to characterize the microwave losses of high-purity aluminum and aluminum alloy under different surface treatments. The results of these measurements and the results of the TEM study are discussed in Sec. \ref{sec:results}.

\section{Loss model in multi-mode cavity resonators} \label{sec:loss model}
The internal quality factors of the resonant modes of a superconducting cavity are dominated by the surface conductive loss of the superconductor, the dielectric loss of the surface oxide, and the seam loss of the joint. Since the resonant modes in the same cavity are affected by the dissipation from the same materials, their quality factors are related to the material loss properties through the matrix equation
\begin{equation}
    \underbrace{\left[ 
    \begin{array}{c} 1/Q_{\mathrm{int}}^{(1)} \\ \vdots \\ 1/Q_{\mathrm{int}}^{(m)} \end{array} 
    \right]}_{\vec{y}}
    = \underbrace{\begin{bmatrix} 1/\mathcal{G}^{(1)} & p_{\mathrm{MA}}^{(1)} & y_{\textrm{seam}}^{(1)} 
    \\ \vdots & \vdots & \vdots 
    \\ 1/\mathcal{G}^{(m)} & p_{\mathrm{MA}}^{(m)} & y_{\textrm{seam}}^{(m)}
    \end{bmatrix}}_{\bm{P}}
    \underbrace{\left[ 
    \begin{array}{c} R_s \\ \tan{\delta} \\ r_{\mathrm{seam}} \end{array} 
    \right]}_{\vec{x}}
    \label{eq:loss_eq}
\end{equation}
where $m$ is the mode index, $\vec{y}$ is a vector containing the reciprocal internal quality factors of the resonant modes, and $\vec{x}$ is a vector containing the material loss factors. These include the surface resistance of the superconductor $R_s$, the scaled loss tangent of the surface oxide $\tan{\delta}$, and the seam resistance per unit length of the joint $r_{\mathrm{seam}}$, which is the reciprocal of the seam conductance per unit length $g_{\mathrm{seam}}$ in the literature \cite{Brecht2015, lei2020high}, i.e., $r_{\mathrm{seam}} = 1/g_{\mathrm{seam}}$. 
$\bm{P}$ is the participation matrix of the system whose rows are the loss participation factors of the resonant modes, which includes the inverse geometric factor\cite{reagor2013reaching, minev2013planar} 
\begin{equation}
    \frac{1}{\mathcal{G}} = 
    \frac{\int_\text{surf}\left|\vec{H}_{\parallel}\right|^2d\sigma}
    {\omega\mu_0\int_\text{vol}\left|\vec{H}\right|^2dv},
\label{eq:1dG}
\end{equation}
the metal-air (MA) surface dielectric participation \cite{reagor2013reaching, wenner2011surface, wang2015surface}
\begin{equation}
    p_{\text{MA}} = 
    \frac{\int_{\text{diel}}\vec{E}\cdot\vec{D}dv}
    {\int_\text{All}\vec{E}\cdot\vec{D}dv} \simeq 
    \frac{t_{\text{MA}}}{\epsilon_r}
    \frac{\int_{\text{surf}}\epsilon_0\left|\vec{E}_{\text{vac}}\right|^2d\sigma}
    {\int_\text{All}\vec{E}\cdot\vec{D}dv},
\label{eq:pMA}
\end{equation}
and the seam admittance per unit length \cite{Brecht2015}
\begin{equation}
    y_{\text{seam}} = 
    \frac{\int_{\text{seam}}\left|\vec{J}_{s}\times\hat{l}\right|^2 dl}
    {\omega\mu_0\int_\text{vol}\left|\vec{H}\right|^2 dv},
\label{eq:yseam}
\end{equation}
where $\vec{H}$, $\vec{E}$, $\vec{D}$, and $\vec{J}_{s}$ are the magnetic field, the electric field, the electric displacement, and the surface current density of the resonant mode. $t_{\text{MA}}$ and $\epsilon_r$ are the thickness and the relative permittivity of the surface oxide. The simplification in Eq. \ref{eq:pMA} is achieved by assuming $t_{\text{MA}}$ is much smaller than the dimension of the cavity and applying the boundary condition $\epsilon E_{\text{diel}, \perp} = \epsilon_0 E_{\text{vac}, \perp}$ at the MA interface, where $\vec{E}_{\text{vac}}$ and $\vec{E}_{\text{diel}}$ are the electric fields in the vacuum region and the surface oxide, respectively.
The loss participation factors can be calculated analytically for simple geometries. For resonators with complicated geometries such as the multi-mode cavity resonators studied in this work, they are calculated numerically using finite-element simulation.

It is important to note that $p_{\mathrm{MA}}$ is proportional to $t_{\mathrm{MA}}$ and inversely proportional to $\epsilon_r$, which are both material properties of the surface oxide. 
In this work, we assume  $t_{\mathrm{MA}}=3\,\textrm{nm}$ and $\epsilon_r = 10$ when calculating $p_{\mathrm{MA}}$, which are typical values for native metal oxides and have been widely used in the literature to calculate the surface dielectric participations \cite{wang2015surface, wenner2011surface, crowley2023disentangling}. 
As a result, the scaled loss tangent $\tan{\delta}$ in E.q.~\ref{eq:loss_eq} is related to the the actual loss tangent of the surface oxide $\tan{\delta_0}$ by 
\begin{equation}
    \tan{\delta} = \frac{t_{\mathrm{MA,0}}}{t_{\mathrm{MA}}} 
                    \frac{\epsilon_{r}}{\epsilon_{r,0}}
                    \tan{\delta_0}.
\label{eq:tand}
\end{equation}
Where $t_{\mathrm{MA},0}$ and $\epsilon_{r,0}$ are the actual thickness and relative dielectric constant of the surface oxide. To determine the actual loss tangent $\tan{\delta_0}$, separate measurement of $t_{\mathrm{MA},0}$ and $\epsilon_{r,0}$ are required.


Since the internal quality factors of the resonant modes are linearly related to the losses associated with the various loss channels, if the rank of the participation matrix is larger than or equal to the number of the loss channels, one can solve Eq.(\ref{eq:loss_eq}) to extract the loss factors using a linear least-squares approach, yielding the solutions:
\begin{equation}
    \vec{x} = \bm{C} \widetilde{\bm{P}}^T \vec{b},
\label{eq:soln}
\end{equation}
where $\vec{b}$ is a vector whose components are $b_i = y_i/\sigma_{y,i}$, $\widetilde{\bm{P}}^T$ is the transpose of the error-weighted participation matrix whose elements are defined as $\widetilde{P}_{ij} = P_{ij}/\sigma_{y,i}$, and $\sigma_{y,i}$ is the standard deviation of $y_i$. The experimental precision determines the relative uncertainty of the measured loss rate, i.e., $\epsilon_{y,i}=\sigma_{y,i}/y_i$. The uncertainties of the extracted material loss factors can be calculated using the covariance matrix
\begin{equation}
    \bm{C}=(\widetilde{\bm{P}}^T \widetilde{\bm{P}})^{-1}.
\label{eq:Cov}
\end{equation}
The diagonal elements of the covariance matrix are the variance of the material loss factors, i.e., $\sigma_{x,i}^2=C_{ii}(\bm{P},\vec{\sigma}_y)$, and the off-diagonal elements describe the correlated errors between the loss channels.
Note that the material loss factors are all positive numbers but the analytic solutions in Eq.(\ref{eq:soln}) do not have this restriction. To add this constraint, we solve Eq.(\ref{eq:loss_eq}) numerically using a non-negative least-squares algorithm and estimate the uncertainties of the material loss factors using a Monte-Carlo analysis \cite{calusine2018analysis, woods2019determining}. 
In this work, both approaches give consistent solutions in most cases; the details of the analysis can be found in the supplementary materials.

Eq.(\ref{eq:soln}) indicates that one can characterize the microwave losses of superconductors by designing particular multi-mode superconducting cavities, measuring the internal quality factors of their resonant modes, and converting them into the material loss factors. In principle, one can solve for the material loss factors $\vec{x}$ using any participation matrix with rank larger than or equal to the number of the loss channels. However, if the rows in the participation matrix are nearly linearly dependent, the solution $\vec{x}$ will be very sensitive to the variations in $\vec{y}$, and the multi-mode system will have very low measurement sensitivity to the material loss factors. In other words, in order to achieve high measurement sensitivity to the material loss factors, the resonant modes in the multi-mode system need to be sensitive to different losses.

An important parameter which can be used to quantify the measurement sensitivity of the system is the relative uncertainty of the material loss factor
\begin{equation}
    \frac{\sigma_{x,i}}{x_i} = \frac{\sqrt{C_{ii}(\bm{P}, \vec{\epsilon}_y, \vec{x})}}{x_i}
    \label{eq:relative_uncertainty}
\end{equation}
For a given participation matrix $\bm{P}$ and relative measurement uncertainty $\vec{\epsilon}_y$, one can estimate the system's measurement sensitivity by evaluating Eq.(\ref{eq:relative_uncertainty}) in the regions of interest in the $\vec{x}$ space. Since the material loss $x_i$ is a positive quantity, the system cannot resolve the material loss $x_i$ when $\sigma_{x,i}/x_i \geq 1$; it can only place an upper bound to the loss. This upper bound is determined by the solutions of $\sqrt{C_{ii}(\bm{P}, \vec{\epsilon}_y, \vec{x})} = x_i$, which divides the $\vec{x}$ space into resolvable and unresolvable regions, determining the measurement sensitivity of the system.

In the following sections, we will present two examples of multi-mode cavity resonators, the forky whispering-gallery-mode resonator (FWGMR) and the ellipsoidal cavity. We use these resonators to characterize microwave losses in superconductors, and apply the formalism introduced in this section to analyze their measurement sensitivity to the material loss factors.

\section{Forky Whispering-Gallery-Mode Resonator (FWGMR)} \label{sec:FWGMR}

\begin{figure}
  \centering
  \includegraphics{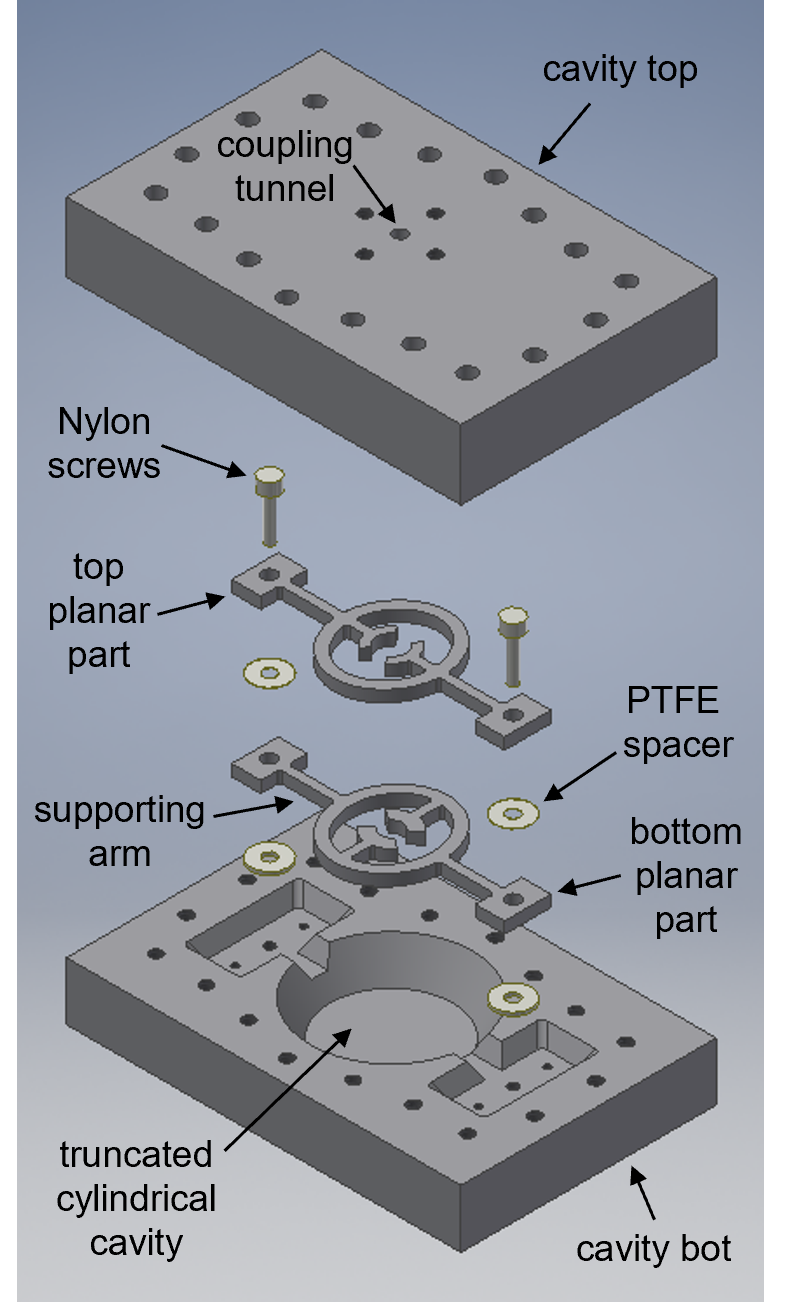}
  \caption{
  Exploded-view diagram of the FWGMR
  }
  \label{fig:FWGMR_drawing}
\end{figure}

\begin{figure*}
  \centering
  \includegraphics{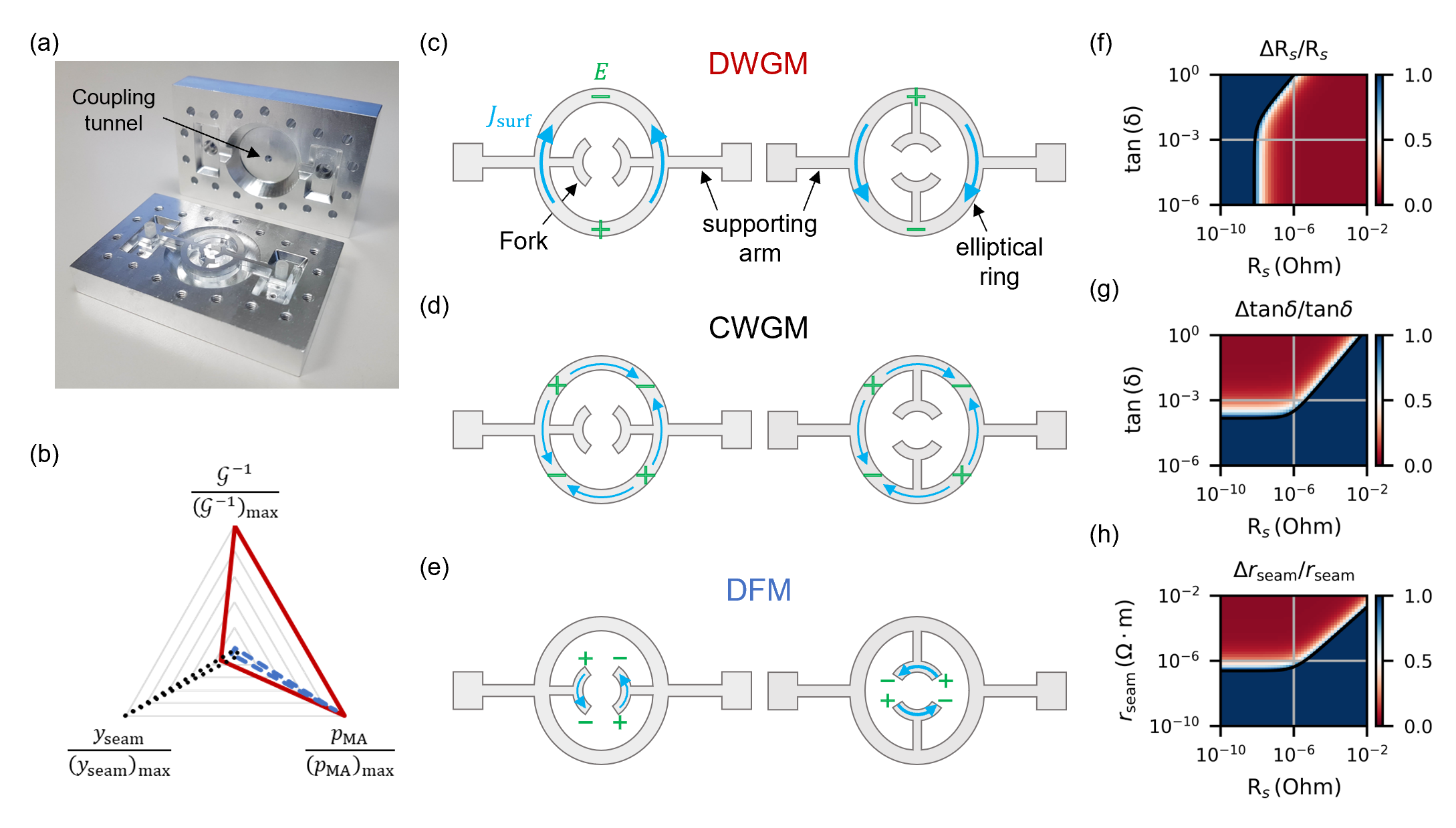}
  \caption{
  (a) Photo of a FWGMR made with high-purity (5N5) aluminum after chemical etching. (b) Normalized participation factors of the differential whispering-gallery mode (DWGM) (red solid line), the differential fork mode (DFM) (dashed blue line), and the common whispering-gallery mode (CWGM) (dotted black line) in a FWGMR with $100\,\upmu\textrm{m}$ gap between the two planar components. The participation factors are normalized by the highest-loss participation factors among the three modes. (c, d, e) Schematic diagrams of the top and bottom planar components, as well as the surface current and electric field configurations of the DWGM (c), the CWGM (d), and the DFM (e). The blue arrows and the green symbols indicate the surface currents and the electric fields. (f, g, h) The sensitivity maps of the multi-mode system formed by the DWGM, the DFM, and the CWGM in the FWGMR with $100\,\upmu\textrm{m}$ gap. (f) and (g) are the sensitivity maps of the surface resistance and the scaled loss tangent at $r_{\text{seam}} = 10^2\,\upmu\Omega\cdot\text{m}$. (h) is the sensitivity map of the seam resistance per unit length at $\tan{\delta} = 5\times10^{-2}$.
  }
  \label{fig:FWGMR1}
\end{figure*}

The first example of multi-mode cavity resonators to characterize microwave losses in superconductors is the forky whispering-gallery-mode resonator. This is a cavity resonator that comprises two planar components that are separated by Teflon spacers, assembled with nylon screws, and enclosed within a truncated cylindrical cavity.  Fig.~\ref{fig:FWGMR_drawing} shows the exploded view of the FWGMR and Fig.~\ref{fig:FWGMR1}a shows the photo of an assembled FWGMR before closing the cavity. Both the planar components and the cavity are made with the superconductor under study using conventional machining processes. The two planar components and the cavity are galvanically isolated from each other by the Teflon spacers. The shapes of the planar components are carefully designed to engineer the frequencies and the loss participation factors of the resonant modes. Each planar part consists of an elliptical ring, a pair of forks connected to the inside of the elliptical ring, and two arms connected to the outside of the ring to provide mechanical support, as shown in Fig.~\ref{fig:FWGMR1}c. The two planar components are separated by approximately $100 \,\upmu\text{m}$ to create resonant modes that are sensitive to the losses from the surfaces. Ideally, the gap size between the planar components is determined by the thickness of the Teflon spacers. In practice, the planar components are not perfectly flat due to imperfections in the machining processes. In addition, thermal contraction during device cool-down can change the parts' dimensions. The gap size could deviate from the thickness of the spacers at room temperature. Here, we infer the gap size using the frequencies of the gap-sensitive modes in the FWGMR (see supplementary materials for details).

The two elliptical rings on the planar components form a whispering-gallery mode resonator (WGMR) \cite{minev2013planar}, which supports the differential whispering-gallery modes (DWGMs) and the common whispering-gallery modes (CWGMs). The DWGMs have opposite charge and current distributions on the two elliptical rings (Fig.~\ref{fig:FWGMR1}c), which confine both the electric and magnetic fields within the vacuum gap between the rings. As a result, these modes are very sensitive to the surface conductive loss of the superconductor and the dielectric loss from the surface oxide, but less sensitive to the seam loss from the cavity joint. 
In contrast, the CWGMs have the same charge and current distributions on the elliptical rings (Fig.~\ref{fig:FWGMR1}d); there are no electromagnetic fields within the vacuum gap. The electromagnetic fields of these modes are more spatially-distributed throughout the cavity and have a much larger mode volume, making them very insensitive to the losses coming from the surfaces but more sensitive to the seam loss from the cavity joint. 

In addition to the elliptical rings, the capacitive coupling between the forks on the planar components generates lumped-element-like resonant modes that we call differential fork modes (DFMs). The frequencies of the DFMs are very sensitive to the gap size between the forks and can be used to infer the distance between the planar components. In these modes, the forks from the top planar component and the forks from the bottom planar component have opposite charge distributions (Fig.~\ref{fig:FWGMR1}e), which concentrates the electric fields within the vacuum gap between the two pairs of forks, making them very sensitive to the dielectric loss from the surface oxide. On the other hand, since the two pairs of forks on the two planar components are oriented orthogonally to each other, the surface currents of these modes do not concentrate magnetic fields within the vacuum gap, making them less sensitive to surface conductive loss as compared to the DWGMs. 

For a FWMGR with $100\, \upmu\text{m}$ gap size, the DWGM, the DFM, and the CWGM form a multi-mode system with participation matrix
\begin{equation} \label{eq:P_FWGMR}
\begin{aligned}
    & \bm{P}_{\text{FWGMR}} = \\
    & \begin{bmatrix} 
    0.28\,(\frac{1}{\Omega}) & 3.8\times10^{-6}  & 
    2.7\times10^{-4}\,(\frac{1}{\Omega\cdot\text{m}})\\[6pt]
    8.9\times10^{-3}\,(\frac{1}{\Omega}) & 3.5\times10^{-6}  & 
    7.1\times10^{-5}\,(\frac{1}{\Omega\cdot\text{m}})\\[6pt]
    5.5\times10^{-3}\,(\frac{1}{\Omega}) & 1.5\times10^{-7} & 
    2.1\times10^{-3}\,(\frac{1}{\Omega\cdot\text{m}})
    \end{bmatrix},
\end{aligned}
\end{equation}
where the columns are the loss mechanisms (inverse geometric factor, surface dielectric participation, and seam admittance per unit length) and the rows are the modes (DWGM, DFM, and CWGM). 

Fig. \ref{fig:FWGMR1}b compares the loss participation factors of the DWGM (solid red line), the DFM (blue dashed line), and the  CWGM (black dotted line), normalized by the highest-loss participation factors among the three modes. It clearly shows that the DWGM and the DFM are much more susceptible to the losses coming from surfaces, whereas the CWGM is more susceptible to the seam loss from the cavity joint. It also shows that the DFM is much less susceptible to surface conductive loss as compared with the DWGM.

Fig.~\ref{fig:FWGMR1}(f, g, h) show the measurement sensitivity of the FWGMR. These sensitivity maps are generated by evaluating the relative uncertainty of the extracted material loss factors (Eq.(\ref{eq:relative_uncertainty})) using the participation matrix $\bm{P}_{\text{FWGMR}}$ (Eq.(\ref{eq:P_FWGMR})) in a selected two-dimensional projection of the material loss space. Here, we use $\epsilon_{y}=5\%$ for all of the modes in the system, as determined by the measurement uncertainty of the internal quality factors. 
Note that a full description of the measurement sensitivity at a point in the material loss space requires three projections for each loss channel; here we only show the three relevant projections for simplicity. A complete set of sensitivity maps can be found in the supplementary materials.

The color in the sensitivity maps represents the relative uncertainty of the material loss factors. Red areas indicate where the material loss is resolvable by the system $(\sigma_{x,i}/x_i <1)$, while blue areas indicate where the material loss is not resolvable by the system $(\sigma_{x,i}/x_i >1)$; only an upper bound to the material loss can be determined. The boundaries between the two regions $(\sigma_{x,i}/x_i =1)$ define the measurement sensitivity of the system, which are indicated by the solid curves in the sensitivity maps.

The measurement sensitivity of a loss channel is not a constant value throughout the material loss space, it depends on the values of the other material loss factors.
Let's take the scaled loss tangent (Fig. \ref{fig:FWGMR1}g) as an example, in the region where $R_s>10^{-6}\,\Omega$, the contour that separates the two regions is a diagonal line. This is because the quality factors of resonant modes are mostly limited by the surface conductive loss in this region; reducing $R_s$ leads to better sensitivity of $\tan{\delta}$. On the other hand, when $R_s \leq 10^{-6}\,\Omega$, the quality factors of the modes are mostly limited by the seam loss; reducing $R_s$ no longer improves the measurement sensitivity of $\tan{\delta}$ and the contour turns into a horizontal line with a minimum resolvable $\tan{\delta}=1.5\times10^{-4}$. Similarly, Fig.~\ref{fig:FWGMR1}f shows the minimum resolvable $R_s = 7.0\,\text{n}\Omega$ for a fixed $r_{\text{seam}}$ and Fig.~\ref{fig:FWGMR1}h shows the minimum resolvable $r_{\text{seam}} = 240\,\text{n}\Omega\cdot\mathrm{m}$ for a fixed $\tan{\delta}$.

In addition to DWGMs, CWGMs, and DFMs, the cavity and the structures providing mechanical support (the supporting arms, the spacers, and the screws) also interact with electromagnetic fields and form their own characteristic resonances, many of which have complicated electromagnetic field distributions. For example,
the galvanic connections between the forks and the elliptical rings provide inductive coupling between them, creating modes that confine the electromagnetic fields within the gap between both the forks and the elliptical rings, which we called the differential forky whispering-gallery modes (DFWGMs). 
The capacitive coupling between the truncated cylindrical cavity and the planar components produces cavity-like resonant modes (CAV) that are less susceptible to the losses from the surface but very sensitive to the seam loss from the cavity joint.
All of these modes can be included in the multi-mode system to extract the material loss factors. The details of the resonant modes in the FWGMR and their loss participation factors can be found in the supplementary materials.

\section{Ellipsoidal cavity}\label{sec:ellipscav}

\begin{figure*}
  \includegraphics{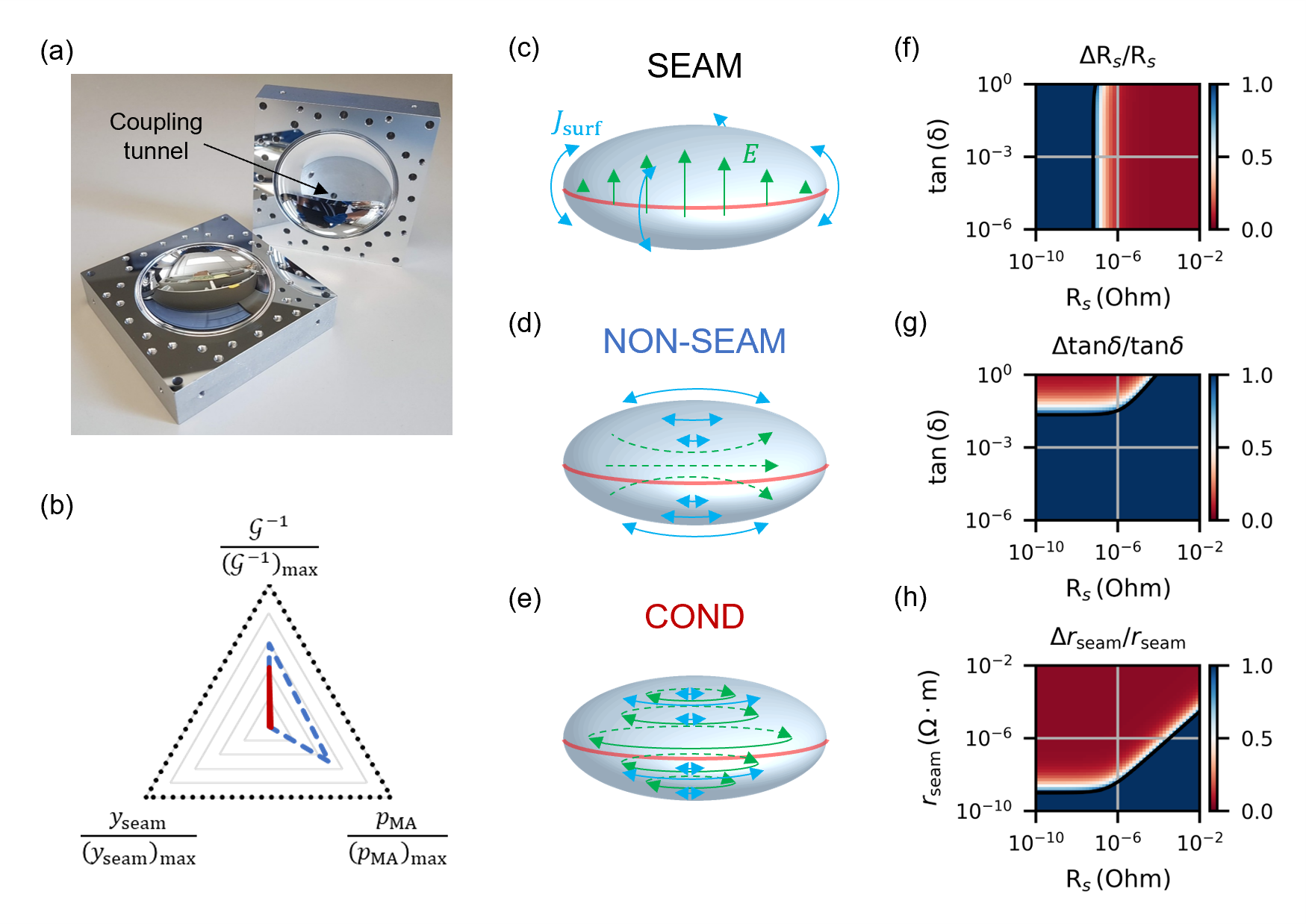}
  \caption{
  (a) Photographs of the two halves of a diamond-turned 6061 aluminum ellipsoidal cavity. (b) Normalized participation factors of the seam loss-sensitive mode (black dotted line), the seam loss-insensitive mode (blue dashed line), and the conductive loss-sensitive mode (red solid line) in the ellipsoidal cavity. The participation factors are normalized by the highest-loss participation factors among the three modes. (c, d, e) Schematic diagrams of the the seam loss-sensitive mode (SEAM), the seam loss-insensitive mode (NON-SEAM), and the conductive loss-sensitive mode (COND). The red solid curves indicate the seam, the blue arrows indicate the surface currents, and the green arrows indicate the electric fields. (f, g, h) The sensitivity maps of the multi-mode system formed by the seam-sensitive mode, the seam-insensitive mode, and the conductive loss-sensitive mode in the ellipsoidal cavity. (f) and (g) are the sensitivity maps of the surface resistance and the scaled loss tangent at $r_{\text{seam}} = 10^2\,\upmu\Omega\cdot\text{m}$. (h) shows the sensitivity maps of the seam resistance per unit length at $\tan{\delta} = 5\times10^{-2}$.
  }
  \label{fig:ellipcav1}
\end{figure*}

While the FWGMR has very high measurement sensitivity to the surface resistance of the superconductor and loss tangent of the surface oxide, it has relatively low sensitivity to the seam resistance of the cavity seam. In addition, its complicated geometry prevents it from measuring the microwave losses of thin-film materials because it is very difficult to homogeneously cover all of the components with thin films without introducing extra interfaces and extra losses. Here, we present another multi-mode superconducting cavity resonator, the ellipsoidal cavity, which is designed to have very high sensitivity to seam resistance and is compatible with thin-film materials.

The ellipsoidal cavity is made of two parts conventionally machined to form halves of an ellipsoid (Fig.~\ref{fig:ellipcav1}a). More specifically, it is an oblate spheroid with major axis equal to 28 mm and minor axis equal to 22.4 mm. The two parts comprising the cavity are bolted together by twenty aluminum screws. The top part contains a coupling tunnel for reflection measurements. Compared to the FWGMR, the ellipsoidal cavity has a simpler geometry and is compatible with surface processing such as diamond turning and thin-film coating.

The modes of the ellipsoidal cavity can be categorized into three types. The first type is the seam loss-sensitive modes (SEAM) (Fig. \ref{fig:ellipcav1}c), e.g., $\text{TM}_{n10}$ with $n\geq0$, etc. The surface currents of these modes are flowing perpendicularly to the seam  and are very sensitive to seam loss from the cavity joint. The second type is the seam loss-insensitive modes (NON-SEAM) (Fig. \ref{fig:ellipcav1}d), e.g., $\text{TE}_{n11}$ with $n\geq1$, etc. In contrast to the seam-sensitive mode, their surface currents flow parallel to and decrease in magnitude towards the seam, making them insensitive to seam loss. The third type is the conductive loss-sensitive modes (COND) (Fig. \ref{fig:ellipcav1}e), e.g., $\text{TE}_{011}$. Similarly to the seam loss-insensitive modes, they are insensitive to seam loss because their surface currents flow parallel to decrease in magnitude towards the seam. Moreover, the electric fields of these modes are azimuthally oriented (Fig. \ref{fig:ellipcav1}e) and thus parallel to the surface of the cavity. As a result, these modes have no electric field on the cavity surface and are insensitive to the surface dielectric losses (Eq. \ref{eq:pMA}). These modes only susceptible to surface conductive loss from the superconductor. Since all modes in the ellipsoidal cavity have a very large mode volume, they are insensitive to the dielectric loss from the surface oxide. 

While the symmetry of the ellipsoidal cavity protects the seam-insensitive modes and the conductive loss-sensitive modes from the seam loss, the coupling tunnel in the cavity or imperfections such as misalignment between the two parts will break the symmetry, perturb the electromagnetic fields, and increase the seam admittance per unit length of the seam-loss insensitive modes and the conductive loss-sensitive modes. To account for these effects, we include the coupling tunnel and consider a $100\,\upmu\text{m}$ offset between the two parts of the cavity in the finite-element simulation when calculating the participation factors. 
The seam loss-sensitive modes, the seam loss-insensitive modes, and the conductive loss-sensitive modes form a multi-mode system with participation matrix
\begin{equation} \label{eq:P_ellip}
\begin{aligned}
    & \bm{P}_{\text{ellip}} = \\
    & \begin{bmatrix} 
    4.3\times10^{-3}\,(\frac{1}{\Omega}) & 3.3\times10^{-8}  & 
    1.3\times10^{-1}\,(\frac{1}{\Omega\cdot\text{m}})\\[6pt]
    2.5\times10^{-3}\,(\frac{1}{\Omega}) & 1.6\times10^{-8}  & 
    5.2\times10^{-5}\,(\frac{1}{\Omega\cdot\text{m}})\\[6pt]
    1.8\times10^{-3}\,(\frac{1}{\Omega}) & 6.7\times10^{-10} & 
    1.6\times10^{-5}\,(\frac{1}{\Omega\cdot\text{m}})
    \end{bmatrix},
\end{aligned}
\end{equation}
where the columns are the loss mechanisms (inverse geometric factor, surface dielectric participation, and seam admittance per unit length) and the rows are the modes (SEAM, NON-SEAM, and COND). 

Compared to the FWMGR, the ellipsoidal cavity has surface dielectric participation factors that are orders of magnitude smaller. Therefore, it is less susceptible to the dielectric loss from the surface oxide. Fig.~\ref{fig:ellipcav1}b compares the normalized participation factors of the seam loss-sensitive modes (black dotted line), the seam loss-insensitive modes (blue dashed line), and the conductive loss-sensitive modes (solid red line). It clearly shows that the seam loss-insensitive modes and the conductive loss-sensitive modes are substantially less susceptible to seam loss than the seam loss-sensitive modes, even considering the imperfection due to the coupling port and the misalignment. It also shows that the conductive loss-sensitive modes are insensitive to the dielectric loss from the surface oxide. 

Fig.~\ref{fig:ellipcav1}(f, g, h) show the measurement sensitivity of the ellipsoidal cavity. Since none of the modes in the ellipsoidal cavity are susceptible to metal-air surface dielectric loss, the measurement sensitivity of the surface resistance is independent of the scaled loss tangent in the region of interest in the material loss space. The minimum resolvable $R_s = 64\,\text{n}\Omega$, at which point the conductive loss-sensitive mode is limited by seam loss due to misalignment between the two halves of the cavity (Fig.~\ref{fig:ellipcav1}f). Compared to the FWGMR, the ellipsoidal cavity has lower sensitivity to the metal-air surface dielectric loss with minimum resolvable $\tan{\delta}=2.2\times10^{-2}$ (Fig.~\ref{fig:ellipcav1}g). On the other hand, it is much more sensitive to the seam loss of the cavity joint. The minimum resolvable $r_{\text{seam}} = 1.0\,\text{n}\Omega\cdot\mathrm{m}$ (Fig.~\ref{fig:ellipcav1}h), which is two orders of magnitude lower than the FWGMR.


\section{Measurements and Results}\label{sec:results}

We use the FWGMR and the ellipsoidal cavity to study the microwave losses of aluminum, as well as the effects of chemical etching, diamond turning, and aluminum thin-film coating on them. The FWGMRs and ellipsoidal cavities studied in this work are made with high-purity (5N5) aluminum and 6061 aluminum alloy using conventional machining processes followed by the surface treatments under study. 
The combinations of the materials and surface treatments of the FWGMRs and the ellipsoidal cavities are shown respectively in TABLE.~\ref{tab:FWGMR_data} and TABLE.~\ref{tab:ellipscav_data}. 

The resonant modes in the FWGMR and the ellipsoidal cavity can be driven by a pin coupler through the coupling tunnel at the top of the cavity (Fig.~\ref{fig:FWGMR1}(a) and Fig.~\ref{fig:ellipcav1}(a)). The coupling strengths to the modes are determined by the location of the coupling tunnel and the length of the coupling pin. They are chosen to be nearly critically coupled or under coupled to the modes of interest. 

The multi-mode resonators are installed in a dilution refrigerator and cooled to 20 mK for measurement (see supplementary materials). We measure the reflection spectra of the resonant modes and extract internal quality factors using the circle fitting method \cite{probst2015efficient}. When measuring the spectra of the differential modes in the FWGMR, i.e., the DWGMs, the DFMs, and the DFWGMs, we observe strong frequency fluctuations due to the electromechanical coupling between the microwave and the mechanical resonances of the planar components, which are driven by the vibration induced by the pulse tube cooler of the dilution refrigerator. To eliminate the frequency fluctuations, we switch off the pulse tube cooler when measuring the spectra of these modes; no appreciable temperature change is observed during the measurement. 

\begin{figure}
  \includegraphics{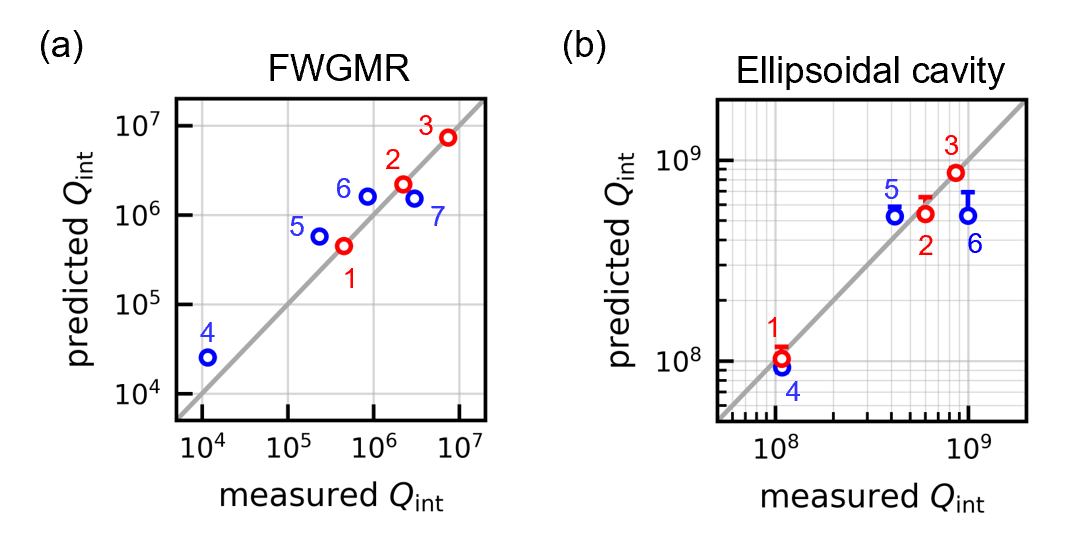}
  \caption{Comparison between the measured internal quality factors and the predicted internal quality factors of the modes. Red (blue) circles are modes that are used (not used) in the material loss analysis. Some of the error bars are smaller than the size of the points and are not visible in these plots. 
  (a) FWGMR (device F4): 1. DWGM-1, 2. DFM-2, 3. CWGM, 4. CAV-2, 5. DWGM-2, 6. DFWGM, 7. DFM-1. (b) ellipsoidal cavity (device E3(eb)): 1. TM310, 2. TE211, 3. TE011, 4. TM210, 5. TE111, 6. TE311.
  }
  \label{fig:QQplots}
\end{figure}

After measuring the internal quality factors, we extract the material loss factors using the methods discussed in section \ref{sec:loss model}. Fig.~\ref{fig:QQplots} compares the predicted internal quality factors with the measured internal quality factors for the modes in the FWGMR (device F4) and the ellipsoidal cavity (device E3(eb)). The diagonal line of slope 1 represents the ideal situation when the predicted values are equal to the measured values. The red circles are the modes used to extract the material loss factors. The blue circles are the modes that are not used in the material loss analysis. 

The predicted internal quality factors are generally consistent with the measured values. However, we observe significant deviations in some of the modes, which could be due to inaccuracies in the participation matrix or other factors that are not taken into account in the loss model. For example, imperfections in the manufacturing and assembly processes could result in deviations from the nominal geometry and induce inaccuracy in the participation matrix. Additionally, the loss model assumes that the material properties are frequency independent over the frequency range of the experiment (3 GHz to 12 GHz) and spatially homogeneous over the devices, which may not be entirely accurate. For example, defects in the raw material or imperfections in the machining processes could potentially lead to spatial variations of the material properties. As a result, resonant modes with different electromagnetic field distributions will experience different average losses. Finally, unknown loss channels that are not included in the loss model could induce additional variations in the predicted quality factors.



\begin{figure*}
  \centering
  \includegraphics{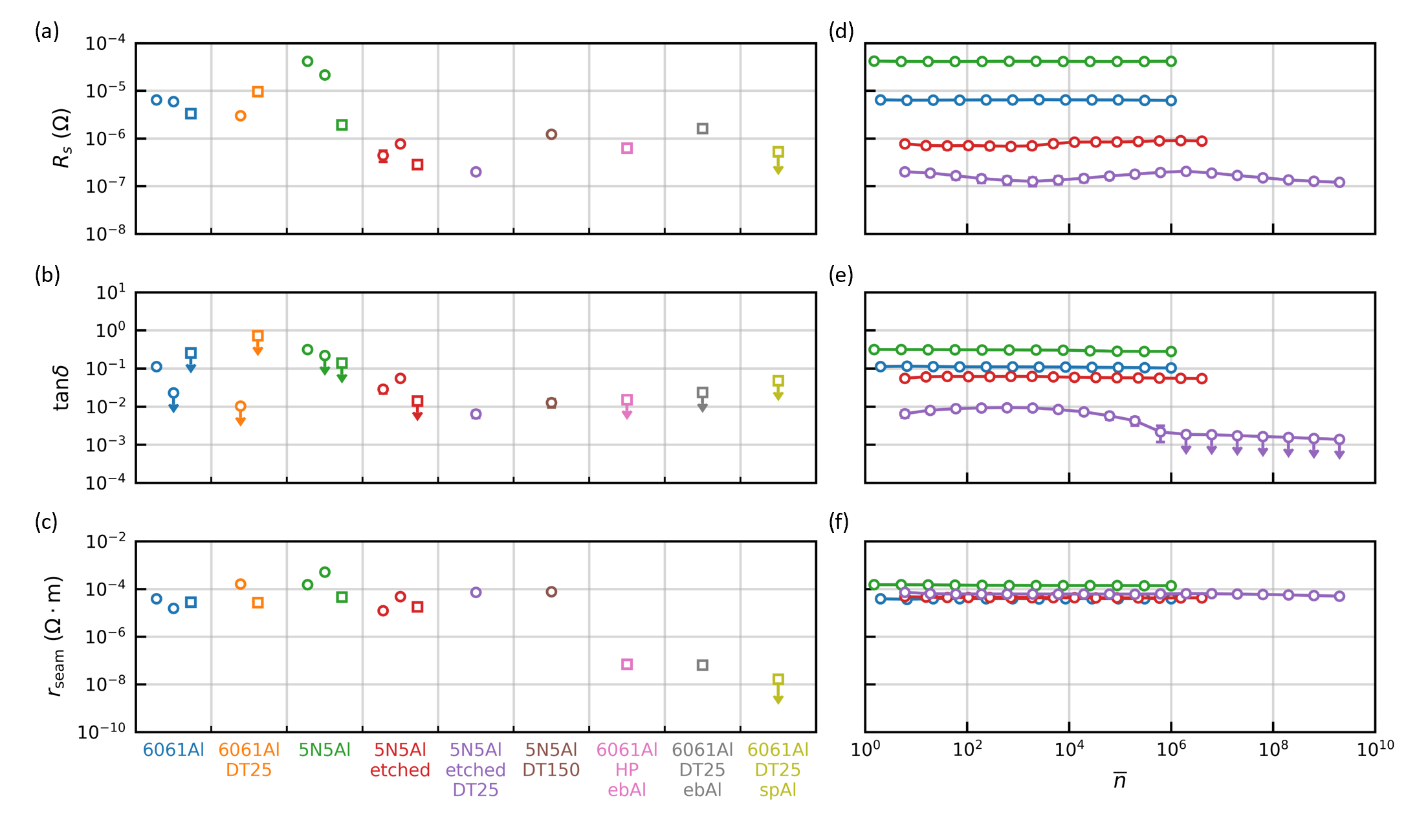}
  \caption{
  Extracted material loss factors and their power dependence. The surface resistance (a), the scaled loss tangent of the surface oxide (b), and the seam resistance per unit length of the joint (c) of untreated aluminum alloy (6061Al), aluminum alloy with 25 $\upmu$m of surface material removed by diamond turning (6061Al DT25), untreated high-purity aluminum (5N5Al), chemically-etched high-purity aluminum (5N5Al etched), chemically-etched high-purity aluminum with 25 $\upmu$m of surface material removed by diamond turning (5N5Al etched DT25), high-purity aluminum with 150 $\upmu$m of surface material removed by diamond turning (5N5Al DT150),
  hand-polished aluminum alloy coated with 600 nm of e-beam evaporated aluminum (6061Al HP ebAl),
  diamond-turned aluminum alloy coated with 600 nm of e-beam evaporated aluminum (6061Al DT25 ebAl), and diamond-turned aluminum alloy coated with 1.6 $\upmu$m of DC magnetron-sputtered aluminum (6061Al DT25 spAl). Circles and squares represent results from the FWGMRs and ellipsoidal cavities respectively. The points with downward arrows represent the upper bounds. (d, e, f) Photon number dependence of the surface resistance (d), the scaled loss tangent of the surface oxide (e), and the seam resistance per unit length (f) extracted from the FWGMRs. The symbol and color scheme are the same as in (a, b, c).
  }
  \label{fig:material_losses_summary}
\end{figure*}

The internal quality factors of the modes involved in the material loss analysis,  their participation factors, and the extracted material loss factors are shown in TABLE.~\ref{tab:FWGMR_data} and TABLE.~\ref{tab:ellipscav_data}, for FWGMRs and ellipsoidal cavities, respectively.
The surface resistance, the scaled loss tangent, and the seam resistance per unit length of the high-purity aluminum and the aluminum alloy with various surface treatments are compared in Fig.~\ref{fig:material_losses_summary}(a-c). Circles and squares represent the material loss factors extracted from FWGMRs and ellipsoidal cavities respectively. Downward arrows indicate the upper bounds of the corresponding material loss factors.

While there is no substantial difference in microwave losses between untreated aluminum alloy (6061Al) and untreated high-purity aluminum (5N5Al), we observe a significant sample-to-sample variation in the surface resistance of the high-purity aluminum devices. The surface resistances extracted from the FWGMRs are an order of magnitude higher than those of the ellipsoidal cavity, which may be due to variations in the quality of the materials or machining processes. 
This discrepancy in material quality is significantly reduced after applying a chemical etching process. The high-purity aluminum devices are etched with Transene aluminum etchant type A at $50\,^{\circ}\text{C}$ for 2 hours to remove approximately 100 $\upmu \text{m}$ of aluminum, followed by a DI water rinse to remove the etchant and then by a blow-dry with nitrogen \cite{reagor2013reaching}.
This process produces aluminum surfaces with more consistent material quality and substantially reduces the surface resistance of the aluminum (5N5Al etched). The average surface resistance of the high-purity aluminum is improved by a factor of 44 and its variation is reduced from $92\%$ to $50\%$. Similar phenomena were also observed in other studies \cite{reagor2013reaching, kudra2020high}.
Additionally, the chemical etching process reduces the scaled loss tangent of the aluminum surface oxide by a factor of 5. Notably, this reduction was only measurable by the FWGMR due to the high sensitivity of the DFM to surface dielectric loss.

To investigate the physical origin of the improvements in material quality due to the chemical etching process, we manufacture two pieces of flat square samples using the same batch of high-purity (5N5) aluminum, similar machining processes, and handling procedures as the high-purity aluminum resonators studied in this work. One of them is chemically etched using the same recipe for etching the high-purity aluminum resonators. We use these samples as proxies for the high-purity aluminum devices and characterize their surfaces using transmission electron microscopy (TEM).


\begin{figure}
  \includegraphics{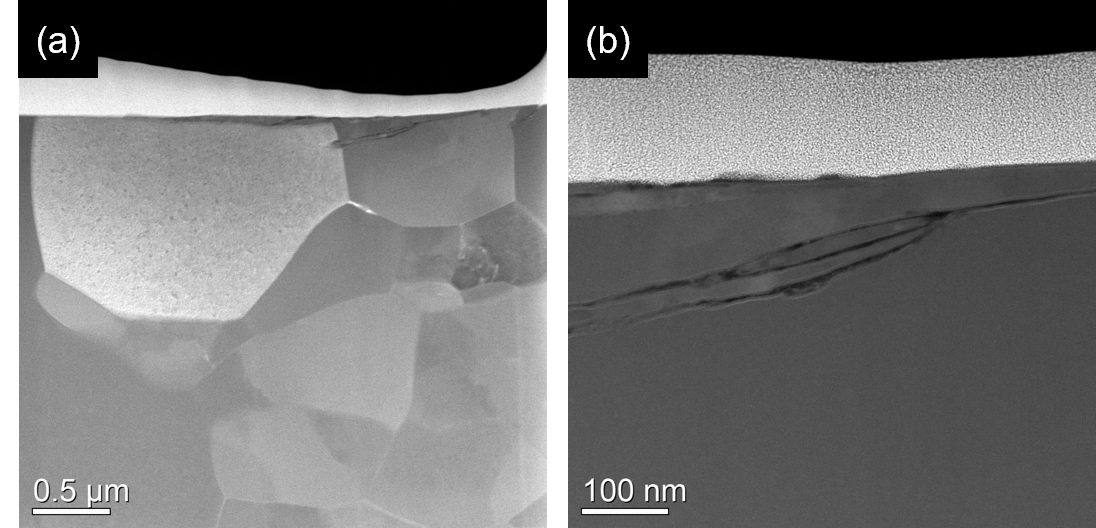}
  \caption{Cross-sectional high-angle annular dark-field scanning TEM (HAADF-STEM) images of an untreated 5N5 aluminum sample. A layer of platinum is deposited on the surface of the sample during the TEM sample preparation. (a) Grains in the bulk aluminum. (b) Subsurface damage in the aluminum.}
  \label{fig:TEM_5N5Al_damage}
\end{figure}

Fig.~\ref{fig:TEM_5N5Al_damage}a shows the cross-section of an untreated 5N5 aluminum sample. The grain size of this sample ranges from several hundred nanometers to a few micrometers. Moreover, we observe damage at about 100 nm below the surface of the sample (Fig.~\ref{fig:TEM_5N5Al_damage}b), which could be generated during the machining processes \cite{mouralova2018quality, zhang2019skin}. 
The observed large fluctuations in the surface resistance of the 5N5 aluminum devices may be coming from the spatial variations in the subsurface damage, which can be removed by the chemical etching process, leading to improved and more consistent material quality. This also implies that other machining or surface processing techniques that can reduce subsurface damage could potentially improve the material quality.

\begin{figure}
  \includegraphics{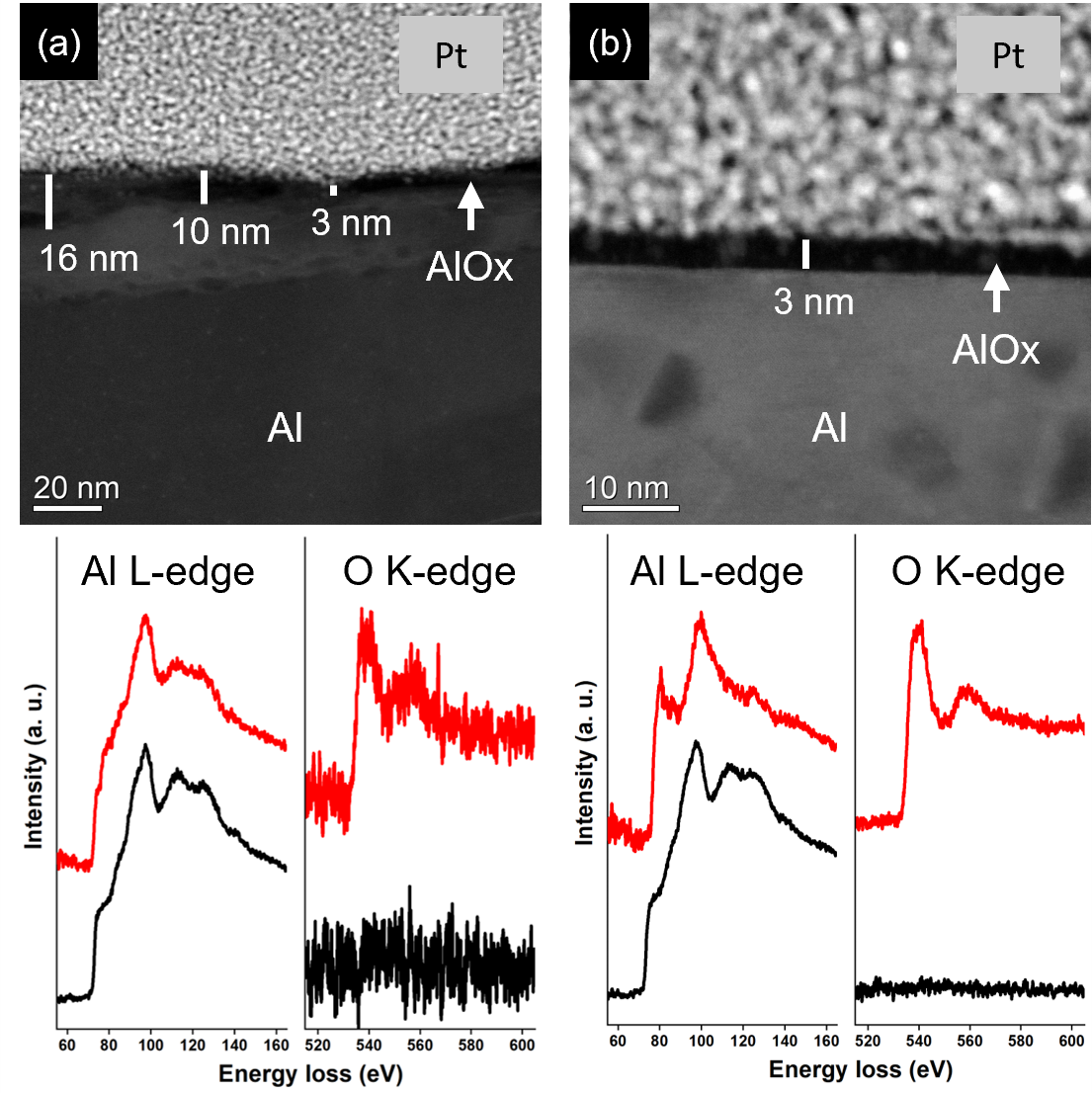}
  \caption{ Cross-sectional HAADF-STEM image of (a) a bare 5N5 aluminum and (b) an etched 5N5 aluminum with Al L-edge and O K-edge EELS spectra from AlOx (red) and Al (black). The grainy material in the upper part of both images is a layer of platinum added during the TEM sample preparation.
  }
  \label{fig:TEM_5N5Al_oxide}
\end{figure}

As discussed in section \ref{sec:loss model}, the reduction in the scaled loss tangent could be caused either by a thinner surface oxide or by improved oxide quality. 
In order to understand the origin of the improvement, an independent measurement of the actual oxide thickness is needed. 
Here, we use TEM to measure the thickness of the aluminum oxide layer on untreated and chemically etched high-purity aluminum samples (Fig.~\ref{fig:TEM_5N5Al_oxide}). Electron energy loss spectra (EELS) were acquired from both AlOx (red) and Al (black), confirming the surface oxidation of the Al. The untreated aluminum surface has an inhomogeneous oxide layer, the oxide thickness ranging from 3 nm to 16 nm with average thickness $t_{\text{MA},0} = 8.82\,\text{nm}$ (Fig.~\ref{fig:TEM_5N5Al_oxide}a), which corresponds to $\tan{\delta_0} = 0.109$. On the other hand, the etched aluminum surface has a uniform oxide layer with thickness  $t_{\text{MA},0} = 3$ nm (Fig.~\ref{fig:TEM_5N5Al_oxide}b), which corresponds to $\tan{\delta_0} = 0.055$. 
These results suggest that the observed improvement in the scaled loss tangent is coming from both the reduction of the oxide thickness and an improvement in the oxide quality. 
Note that the scaled loss tangent measured in this work is an average value over the cavity surface, whereas the oxide thickness measured from the TEM image is localized around a small area of the sample. To quantify the variation in the surface oxide's loss tangent, further confirmation of the oxide thickness on a large scale is necessary.

As suggested from the TEM study, improving the surface quality of the material could potentially reduce its microwave losses and result in more consistent material loss properties.
Besides chemical etching, an alternative method to improve the surface quality of materials is diamond turning, which is a precision machining process to uniformly remove material using a precision CNC lathe and a diamond-tipped cutting tool. It is a chemical-free process that can create mirror-finished surfaces with consistent surface quality (average roughness on the order of 10 nm). Here, we diamond turn the surfaces of the multi-mode resonators to study its effect on the microwave loss of aluminum. 
The average roughness of the diamond-turned aluminum surfaces is around 20 nm, which is about 50 times smaller than the untreated aluminum surfaces (see supplementary materials). Fig.~\ref{fig:ellipcav1}a shows an aluminum ellipsoidal cavity after diamond turning.

Although there is no improvement in aluminum alloy after removing $25\,\upmu\text{m}$ of material with diamond turning (6061Al DT25), the microwave losses of the high-purity aluminum are significantly improved after a $150\,\upmu\text{m}$ diamond turn (5N5Al DT150) (Fig. \ref{fig:material_losses_summary}). Its surface resistance and scaled loss tangent are reduced to roughly the same level as the chemically etched high-purity aluminum, indicating that diamond turning may be a chemical-free alternative to etching high-purity aluminum.
To evaluate the effects of combining chemical etching with diamond turning, we performed a 25 $\upmu\text{m}$ diamond turn of the chemically etched high-purity aluminum FWGMR. The surface resistance is improved by a factor of 2.2, and the scaled loss tangent is improved by a factor of 4.4 (5N5Al etched DT25) after diamond turning. 

\begin{figure}
  \includegraphics{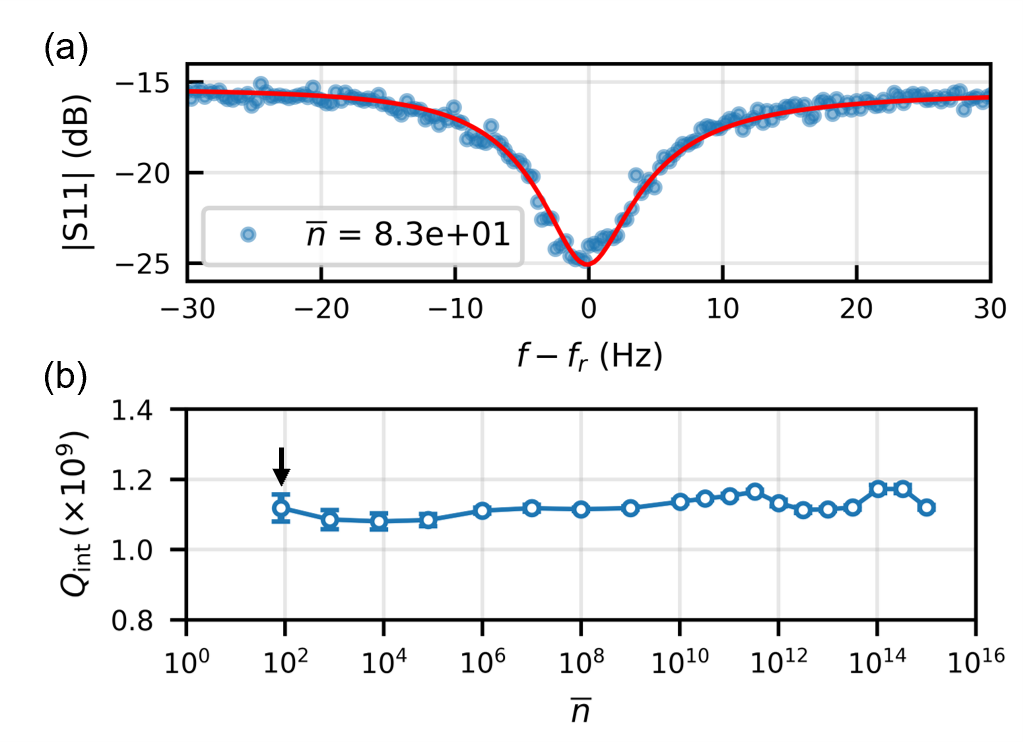}
  \caption{Reflection spectrum of the the $\text{TM}_{311}$ mode of the hand-polished ellipsoidal cavity coated in e-beam-evaporated aluminum (device E3(eb)) (a) and its internal quality factor as a function of average photon number (b). The red curve in (a) is the result of the circle fit, its internal quality factor is indicated by the black arrow in (b).}
  \label{fig:mode15_S21_n_nbar_sweep}
\end{figure}

In addition to diamond turning, thin-film coating is another method to create high quality superconducting surfaces \cite{kuhr2007ultrahigh, benvenuti1999study, roach2012niobium}. Here, we use ellipsoidal cavities to characterize the microwave losses of aluminum thin-film deposited on polished aluminum alloy surfaces and diamond-turned aluminum alloy surfaces. Prior to thin-film coating, The cavities are either hand polished (HP) (device E3) or diamond turned (DT) (device E4) to improve the average surface roughness to the level of a few tens of nanometers before thin-film coating. They are coated with 600 nm of e-beam evaporated aluminum (6061Al HP ebAl and 6061Al DT25 ebAl). Following the first measurement, device E4 is coated with 1.6 $\upmu\text{m}$ of magnetron-sputtered aluminum on top of the 600 nm of e-beam evaporated aluminum (6061Al DT spAl). 

The thin-film coating process improves the microwave losses of all three ellipsoidal cavities. Their surface resistances are reduced to the same level as the chemically-etched high-purity aluminum. Their scaled loss tangents are improved below the system's measurement sensitivity, with upper bounds at the same level as the chemically etched high-purity aluminum. Additionally, thin film coating produces lower seam resistances than any other surface treatment by over 2 orders of magnitude. In particular, the sputter-coating method produces seam resistances lower than the measurement sensitivity of the system.
As a result, the internal quality factors of the modes in these devices are very high due to the small loss participation factors and the improved microwave losses. The internal quality factor of the seam loss-insensitive mode ($\text{TE}_{311}$) of device E3(eb) reaches above one billion without appreciable power dependence (Fig.~\ref{fig:mode15_S21_n_nbar_sweep}), which is as good as the conductive loss-sensitive mode ($\text{TE}_{011}$) of the chemically-etched aluminum ellipsoidal cavity. The thin-film coating method therefore provides a path towards achieving high coherence without the need for bulk high-purity aluminum, which is both costly and difficult to machine.

Finally, by extracting the material loss factors as a function of circulating photon number, we can study the power dependence of the microwave losses in the FWGMRs as shown in Fig.~\ref{fig:material_losses_summary} (d-f). 
While the surface resistance and the seam resistance per unit length show no appreciable power dependence for all of the devices,
the scaled loss tangent of the diamond-turned chemically-etched high-purity aluminum (purple circles) decreases when the average photon number increases above the critical number $n_c=10^4$, and continues to decrease beyond the system's measurement sensitivity at $(\bar{n}\geq10^6)$. This indicates that the microwave loss of its surface oxide is limited by two-level systems (TLS) with a critical electric field $E_{c} \simeq 3\, \text{V/m}$, which is of the same order of magnitude as the critical electric field observed in \cite{kudra2020high} ($E_{c} \simeq 2\, \text{V/m}$). On the other hand, no appreciable power dependence in the scaled loss tangent is observed in the other devices up to $\bar{n} \simeq 10^6$, which clearly shows that diamond-turning modifies the properties of the surface oxide on the high-purity aluminum.

\begin{figure}
  \includegraphics{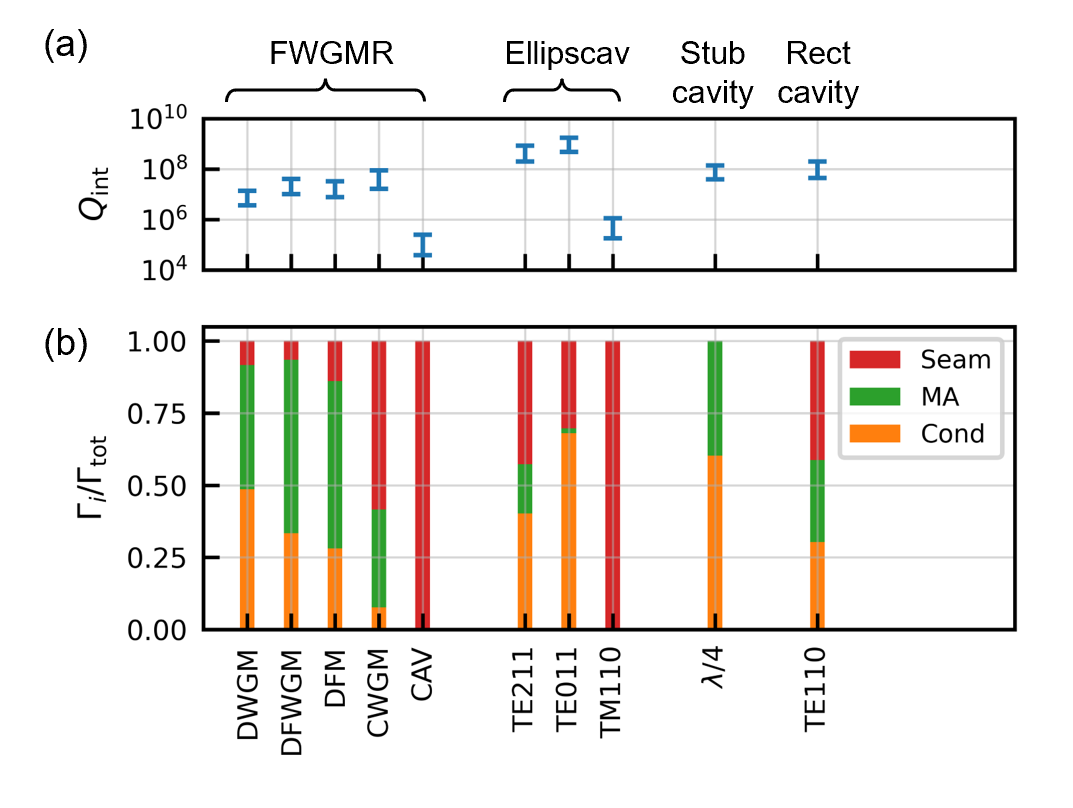}
  \caption{(a) The predicted quality factors of the resonant modes in the FWGMR, ellipsoidal cavity (Ellipscav), coaxial stub cavity (Stub cavity), and rectangular cavity (Rect cavity) using the average surface resistance, scaled loss tangent, and seam resistance per unit length of the chemically-etched high-purity aluminum, i.e., $R_s = 500\pm250\,\text{n}\Omega$, $\tan{\delta} = 0.033\pm0.021$, and $r_{\text{seam}} = 26\pm19\, \upmu\Omega\cdot\text{m}$. (b) The relative contributions of the surface conductive loss (Cond), the surface dielectric loss (MA), and the seam loss (Seam) of the modes in (a).}
  \label{fig:predicted_q_n_losses}
\end{figure}

The extracted material loss factors can be used to estimate the loss contributions from the various loss channels and predict the internal quality factors of other device made with the same materials and fabrication processes. 
Here, we analyze the losses in chemically-etched high-purity aluminum superconducting cavities (Fig.~\ref{fig:predicted_q_n_losses}), including the multi-mode cavities studied in this work, as well as the coaxial stub cavity \cite{Reagor2016, kudra2020high} and the rectangular cavity \cite{reagor2013reaching, Brecht2015}, which are widely used as quantum memories in superconducting quantum circuits \cite{ofek2016extending, chou2018deterministic, zhou2021modular}. Note that we are considering $100\,\upmu\text{m}$ of misalignment between the two halves of the cavities when calculating their seam loss.
The predicted internal quality factors are consistent with the experimental observations in this work and other studies \cite{reagor2013reaching, Brecht2015, kudra2020high, read2022precision, Reagor2016} (Fig.~\ref{fig:predicted_q_n_losses}a). 
The fractional contributions of the surface conductive loss (Cond), the surface dielectric loss (MA), and the seam loss (Seam) of these devices are shown in Fig.~\ref{fig:predicted_q_n_losses}b. 

As discussed in Sec. \ref{sec:FWGMR}, the differential modes (DWGM, DFWGM, DFM) in the FWGMR are insensitive to seam loss, with less than $20\%$ of their loss coming from the seam (Fig.~\ref{fig:predicted_q_n_losses}b). In particular, the DFM and the DFWGM have more than $50\%$ of the loss from the MA interface. On the other hand, the modes in the ellipsoidal cavity only have less than $20\%$ of loss coming from the MA interface, which is consistent with the higher measurement sensitivity to the scaled loss tangent in the FWGMR. 


Although the quarter-wave mode in the coaxial stub cavity and the $\text{TE}_{110}$ mode in the rectangular cavity have similar internal quality factors (Fig.~\ref{fig:predicted_q_n_losses}a), the origins of their losses are different. 
The quarter-wave mode is insensitive to seam loss: 60\% of the loss comes from the residual resistance in the superconductor, and 40\% of the loss comes from the microwave absorption in the surface oxide. 
In contrast, about $41\%$ of the loss in the $\text{TE}_{110}$ mode comes from the seam, $30\%$ comes from the residual resistance in the superconductor, and only $29\%$ comes from the microwave absorption in the surface oxide. This indicates that the quality factor of the $\text{TE}_{110}$ mode can be further improved if one can eliminate the loss from the seam, which can be achieved by improving the seam quality \cite{lei2020high, romanenko2020three}, or creating the cavity with a seamless architecture \cite{chakram2021seamless}. 

It is important to note that even though we operate the aluminum superconducting cavities at 20 mK, which is significantly lower than its superconducting transition temperature, we still observed a significant amount of surface conductive loss from the superconductor. The corresponding surface resistance is much higher than the prediction from the BCS theory,  and the origin of the observed residual surface resistance remains unknown. For the chemically etched high-purity aluminum studied in this work, the average surface resistance is $500\pm250\,\text{n}\Omega$, which is higher than the surface resistance of aluminum thin film on crystallized substrates that is typically below $250\,\text{n}\Omega$ \cite{minev2013planar}. 

In addition to the superior surface resistance, thin film superconductors can also create joints with very low seam loss \cite{Brecht2015, lei2020high}. This suggests that constructing 3D superconducting resonators with crystalline substrate and superconducting thin film can further improve the quality factors by reducing the material loss factors. This approach has been demonstrated using evaporated indium thin films and micromachined silicon substrates \cite{Brecht2015, lei2020high}. 

Another promising approach for improving the coherence of superconducting resonators is to create hybrid 3D superconducting resonators that combine traditional or micromachining-based 3D superconducting enclosures with on-chip components made with crystalline substrates and superconducting thin film. The on-chip components not only have superior material quality than bulk superconductors, they can be used to engineer the electromagnetic field distribution of the resonant modes. Careful design of the hybrid structure can lead to optimized loss participation factors of both the enclosure and the on-chip components, resulting in significant improvements in the device coherence \cite{Krayzman2023, Ganjam2023}.

\section{Conclusions}
We have introduced a method to measure microwave losses of materials using multi-mode resonators. We have presented two types of multi-mode superconducting cavity resonators for the study of microwave losses in bulk superconductors: the FWGMR, and the ellipsoidal cavity. These losses include the surface resistance of the superconductor, the scaled loss tangent of the surface oxide, and the seam resistance per unit length of the joint. We have used these multi-mode resonators to measure the microwave losses of aluminum alloy and high-purity aluminum, as well as quantify the effects of chemical etching, diamond turning, and aluminum thin-film coating on the microwave losses. 

We have found that the chemical etching process not only improves the surface resistance of the high-purity aluminum, but also reduces the scaled loss tangent of its surface oxide. We have studied high-purity aluminum and the chemical etching process using transmission electron microscopy (TEM). We observed subsurface damage in the aluminum and varying thickness of surface oxide in the untreated aluminum sample, whereas the surface of the chemically etched aluminum sample had a more uniform surface oxide layer. Besides chemical etching, we also studied the effect of diamond turning. While no improvement in the microwave losses is observed in aluminum alloy, diamond-turning high-purity aluminum lowers the surface resistance and scaled loss tangent to levels similar to chemical etching. Moreover, we found that the scaled loss tangent of the chemically-etched high-purity aluminum improves with increased average cavity photon number after diamond turning, which indicates that its surface dielectric loss is limited by two-level systems (TLS) and the diamond turning process modifies the TLS properties of the surface oxide.

In addition, the internal quality factor of the $\text{TM}_{311}$ mode in the aluminum alloy ellipsoidal cavity improves above one billion after its surfaces are polished and coated with e-beam evaporated aluminum thin film. We found that coating polished or diamond-turned surfaces with aluminum thin films not only improves the surface resistance and the scaled loss tangent, but also significantly reduces the seam resistance per unit length by more than three orders of magnitude. 
Unlike the chemical etching process, which would reduce the precision of the parts' dimensions, diamond turning and thin-film coating can produce parts with very high precision and excellent surface finishing. More importantly, being able to create very high-quality seams enables the realization of superconducting cavities and enclosures with more complicated geometries, which is critical to scaling up cavity-based superconducting quantum devices \cite{Brecht2015, lei2020high}.

Finally, the methods introduced in this paper are not limited to studying microwave losses in bulk superconductors. Similar concepts can be applied to design on-chip superconducting devices to measure microwave losses in superconducting thin films and substrate materials, as well as quantify the effects of fabrication processes, which are crucial to the development of high-coherence superconducting quantum devices.

\section{Acknowledgements}
We thank Benjamin Chapman, Alexander Read, and Ignace Jarrige for useful discussions; 
Yong Sun, Sean Rinehart, and Kelly Woods for assistance with device fabrication. 
This research was supported by the U.S. Army Research Office (ARO) under grant W911NF-18-1-0212. The views and conclusions contained in this document are those of the authors and should not be interpreted as representing official policies, either expressed or implied, of the ARO or the U.S. Government. The U.S. Government is authorized to reproduce and distribute reprints for Government purpose notwithstanding any copyright notation herein. Fabrication facilities use was supported by the Yale Institute for Nanoscience and Quantum Engineering (YINQE) and the Yale SEAS Cleanroom. This research used electron microscopy facility of the Center for Functional Nanomaterials (CFN), which is a U.S. Department of Energy Office of Science User Facility, at Brookhaven National Laboratory under Contract No. DE-SC0012704.
L.F. and R.J.S. are founders and shareholders of Quantum Circuits, Inc.

\begin{table*}[b]
\caption{
\label{tab:FWGMR_data}
Data from the FWGMRs.
5N5Al: high-purity ($99.9995\%$) aluminum. 
6061Al: 6061 aluminum alloy. 
etched: chemical etching with aluminum etchant type A at $50\,^{\circ}\text{C}$ for 2 hours. 
DT25: Diamond turned 25 $\upmu\text{m}$ of cavity surfaces. 
DT150: Diamond turned 150 $\upmu\text{m}$ depth of cavity surfaces.
Gap indicates the distance between the two planar components, which is determined by the frequencies of the modes (see supplementary materials).
}
\begin{ruledtabular}
\begin{tabular}{cccccccccc}
 \thead{Device\\(gap)}&
 Material&
 Mode&
 \thead{Freq\\(GHz)}&
 \thead{$Q_c$\\$(\times10^6)$}&
 \thead{$Q_{\textrm{int}}$\\$(\times10^6)$}&
 \thead{$1/\mathcal{G}$\\$(1/\Omega)$}&
 \thead{$p_{\textrm{MA}}$}&
 \thead{$y_{\textrm{seam}}$\\$(/\Omega/\textrm{m})$}&
 \thead{
 $R_s\,(\upmu\Omega)$\\
 $\tan{\delta}$\\
 $r_{\textrm{seam}}\,(\upmu\Omega\cdot\textrm{m})$
 }
 \\
\hline
\thead{F1\\(65 $\upmu$m)}& 5N5Al& 
\thead{DWGM-1\\DFM-2\\CWGM-1}&
\thead{5.590\\7.997\\10.862}&
\thead{$3.2$\\$0.36$\\$6.4$}&
\thead{$0.047$\\$0.43$\\$1.74$}&
\thead{$0.46$\\$1.1\times10^{-2}$\\$5.5\times10^{-3}$}&
\thead{$6.4\times10^{-6}$\\$5.9\times10^{-6}$\\$1.5\times10^{-7}$}&
\thead{$2.4\times10^{-4}$\\$8.0\times10^{-5}$\\$2.0\times10^{-3}$}&
\thead{$41.8\pm2.4$\\$0.32\pm0.02$\\$153\pm16$}\\ 
\hline
\thead{F1(e)\\(80 $\upmu$m)}&\thead{5N5Al\\etched}& 
\thead{DWGM-1\\DFWGM-1\\CWGM-1}&
\thead{5.756\\6.465\\10.879}&
\thead{$7.0$\\$1.5$\\$12$}&
\thead{$3.2$\\$12$\\$59$}&
\thead{$0.37$\\$6.9\times10^{-2}$\\$5.6\times10^{-3}$}&
\thead{$5.1\times10^{-6}$\\$1.8\times10^{-6}$\\$1.6\times10^{-7}$}&
\thead{$4.0\times10^{-4}$\\$2.2\times10^{-4}$\\$8.0\times10^{-4}$}&
\thead{$0.44\pm0.11$\\$0.029\pm0.006$\\$12.5\pm1.3$}\\ 
\hline
\thead{F2\\(155 $\upmu$m)}&5N5Al& 
\thead{DWGM-1\\DFM-1\\CWGM-1}&
\thead{6.043\\3.696\\10.873}&
\thead{$0.44$\\$1.6$\\$5.7$}&
\thead{$0.17$\\$0.96$\\$0.88$}&
\thead{$0.16$\\$4.5\times10^{-2}$\\$5.4\times10^{-3}$}&
\thead{$2.2\times10^{-6}$\\$1.5\times10^{-6}$\\$2.0\times10^{-7}$}&
\thead{$4.0\times10^{-4}$\\$6.3\times10^{-4}$\\$1.8\times10^{-3}$}&
\thead{$21.5\pm1.1$\\$\leq0.22$\\$512\pm32$}\\ 
\hline
\thead{F2(e)\\(150 $\upmu$m)}&\thead{5N5Al\\etched}& 
\thead{DWGM-1\\DFM-2\\CAV-1}&
\thead{6.040\\10.562\\8.698}&
\thead{$4.1$\\$1.1\times10^2$\\$1.1\times10^{-2}$}&
\thead{$3.5$\\$8.7$\\$3.6\times10^{-2}$}&
\thead{$0.17$\\$7.0\times10^{-3}$\\$6.2\times10^{-3}$}&
\thead{$2.3\times10^{-6}$\\$1.9\times10^{-6}$\\$2.3\times10^{-7}$}&
\thead{$5.7\times10^{-4}$\\$1.2\times10^{-4}$\\$0.58$}&
\thead{$0.78\pm0.10$\\$0.055\pm0.003$\\$47.8\pm2.4$}\\ 
\hline
\thead{F2(ed)\\(68 $\upmu$m)}&\thead{5N5Al\\etched\\DT25}& 
\thead{DWGM-1\\DFM-1\\CWGM-1}&
\thead{5.790\\3.208\\10.881}&
\thead{$5.5$\\$9.0$\\$37$}&
\thead{$6.0$\\$13$\\$19$}&
\thead{$0.46$\\$5.8\times10^{-2}$\\$5.3\times10^{-3}$}&
\thead{$6.2\times10^{-6}$\\$4.1\times10^{-6}$\\$1.4\times10^{-7}$}&
\thead{$4.7\times10^{-4}$\\$5.4\times10^{-4}$\\$6.9\times10^{-4}$}&
\thead{$0.20\pm0.03$\\$0.0065\pm0.0013$\\$72.8\pm3.9$}\\ 
\hline
\thead{F5(d)\\(88 $\upmu$m)}&\thead{5N5Al\\DT150}& 
\thead{DWGM-1\\DFM-2\\CAV-1}&
\thead{5.734\\3.281\\8.631}&
\thead{$3.3$\\$23$\\$7.8\times10^{-3}$}&
\thead{$2.1$\\$5.0$\\$2.4\times10^{-2}$}&
\thead{$0.34$\\$5.3\times10^{-2}$\\$6.1\times10^{-3}$}&
\thead{$4.6\times10^{-6}$\\$3.3\times10^{-6}$\\$2.4\times10^{-7}$}&
\thead{$1.4\times10^{-4}$\\$3.4\times10^{-4}$\\$0.28$}&
\thead{$0.95\pm0.11$\\$0.030\pm0.004$\\$152\pm8$}\\ 
\hline
\thead{F3\\(100 $\upmu$m)}&\thead{6061Al}& 
\thead{DWGM-1\\DFM-1\\
CAV-2\footnote{
To account for the additional uncertainty from the bulk dielectric loss of the screws and the washers, the relative uncertainty of the measured loss rate $\epsilon_y$ is set to $20\%$ for this mode (see supplementary materials).}
}&
\thead{5.802\\3.417\\11.081}&
\thead{$7.3$\\$15$\\$2.1\times10^{-2}$}&
\thead{$0.56$\\$3.4$\\$6.5\times10^{-2}$}&
\thead{$0.28$\\$5.1\times10^{-2}$\\$3.8\times10^{-6}$}&
\thead{$3.8\times10^{-6}$\\$2.7\times10^{-6}$\\$8.2\times10^{-7}$}&
\thead{$2.7\times10^{-4}$\\$6.4\times10^{-4}$\\$1.0$}&
\thead{$5.96\pm0.21$\\$\leq0.023$\\$15.4\pm3.6$}\\ 
\hline
\thead{F3(d)\\(72 $\upmu$m)}&\thead{6061Al\\DT25}& 
\thead{DWGM-1\\DFM-1\\CWGM-1}&
\thead{5.784\\3.183\\10.865}&
\thead{$0.86$\\$3.0$\\$26$}&
\thead{$0.77$\\$3.67$\\$7.97$}&
\thead{$0.42$\\$5.7\times10^{-2}$\\$5.8\times10^{-3}$}&
\thead{$5.7\times10^{-6}$\\$4.2\times10^{-6}$\\$1.6\times10^{-7}$}&
\thead{$1.7\times10^{-4}$\\$5.7\times10^{-4}$\\$6.6\times10^{-4}$}&
\thead{$3.01\pm0.17$\\$\leq0.010$\\$162\pm10$}\\ 
\hline
\thead{F4\\(100 $\upmu$m)}&\thead{6061Al}& 
\thead{DWGM-1\\DFM-2\\CWGM-1}&
\thead{5.858\\9.199\\10.863}&
\thead{$1.6$\\$14$\\$5.7$}&
\thead{$0.45$\\$2.2$\\$7.4$}&
\thead{$0.28$\\$8.9\times10^{-3}$\\$5.5\times10^{-3}$}&
\thead{$3.8\times10^{-6}$\\$3.5\times10^{-6}$\\$1.5\times10^{-7}$}&
\thead{$2.7\times10^{-4}$\\$7.1\times10^{-5}$\\$2.1\times10^{-3}$}&
\thead{$6.48\pm0.43$\\$0.11\pm0.01$\\$39.1\pm3.5$}\\ 
\end{tabular}
\end{ruledtabular}
\end{table*}

\begin{table*}[b]
\caption{
\label{tab:ellipscav_data}
Data from the ellipsoidal cavities. 
5N5Al: high-purity ($99.9995\%$) aluminum. 
6061Al: 6061 aluminum alloy. 
etched: chemical etching with aluminum etchant type A at $50\,^{\circ}\text{C}$ for 2 hours. 
HP: Hand polished with sandpaper and finished with alumina-based metal polishing compound (Pikal Care). 
DT25: Diamond turned 25 $\upmu$m of cavity surfaces. 
ebAl: cavity surfaces coated with e-beam evaporated 600 nm aluminum thin-film.
spAl: cavity surfaces coated with with magnetron-sputtered 1.6 $\upmu\text{m}$ aluminum thin film.
}
\begin{ruledtabular}
\begin{tabular}{cccccccccc}
 Device&
 Material&
 Mode&
 \thead{Freq\\(GHz)}&
 \thead{$Q_c$\\$(\times10^6)$}&
 \thead{$Q_{\textrm{int}}$\\$(\times10^6)$}&
 \thead{$1/\mathcal{G}$\\$(1/\Omega)$}&
 \thead{$p_{\textrm{MA}}$}&
 \thead{$y_{\textrm{seam}}$\\$(/\Omega/\textrm{m})$}&
 \thead{
 $R_s\,(\upmu\Omega)$\\
 $\tan{\delta}$\\
 $r_{\textrm{seam}}\,(\upmu\Omega\cdot\textrm{m})$
 }
 \\
\hline
E1& 5N5Al&
\thead{$\text{TM}_{310}$\\$\text{TE}_{111}$\\$\text{TE}_{011}$}&
\thead{11.556\\8.450\\10.723}&
\thead{$0.12$\\$46$\\$2.4\times10^4$}&
\thead{$0.21$\\$183$\\$195$}&
\thead{$3.0\times10^{-3}$\\$2.8\times10^{-3}$\\$1.8\times10^{-3}$}&
\thead{$4.3\times10^{-8}$\\$0.8\times10^{-8}$\\$6.7\times10^{-10}$}&
\thead{$0.10$\\$1.5\times10^{-5}$\\$1.6\times10^{-5}$}&
\thead{$1.93\pm0.08$\\$\leq0.14$\\$45.9\pm2.3$}\\
\hline
E1(e)& \thead{5N5Al\\etched}&
\thead{$\text{TM}_{110}$\\$\text{TE}_{211}$\\$\text{TE}_{011}$}&
\thead{7.225\\10.216\\10.731}&
\thead{$0.44$\\$92$\\$7.0\times10^3$}&
\thead{$0.43$\\$644$\\$1.2\times10^3$}&
\thead{$4.3\times10^{-3}$\\$2.5\times10^{-3}$\\$1.8\times10^{-3}$}&
\thead{$3.3\times10^{-8}$\\$1.6\times10^{-8}$\\$6.7\times10^{-10}$}&
\thead{$0.13$\\$5.2\times10^{-5}$\\$1.6\times10^{-5}$}&
\thead{$0.29\pm0.02$\\$\leq0.014$\\$18.0\pm0.90$}\\
\hline
E2& \thead{6061Al}&
\thead{$\text{TM}_{020}$\\$\text{TE}_{111}$\\$\text{TE}_{011}$}&
\thead{10.001\\8.479\\10.756}&
\thead{$0.1$\\$37$\\$1.1\times10^3$}&
\thead{$0.29$\\$91$\\$150$}&
\thead{$3.3\times10^{-3}$\\$2.8\times10^{-3}$\\$1.8\times10^{-3}$}&
\thead{$3.9\times10^{-8}$\\$0.8\times10^{-8}$\\$6.7\times10^{-10}$}&
\thead{$0.12$\\$1.5\times10^{-5}$\\$1.6\times10^{-5}$}&
\thead{$3.32\pm0.20$\\$\leq0.26$\\$28.0\pm1.4$}\\
\hline
E3(eb)& \thead{6061Al\\HP\\ebAl}&
\thead{$\text{TM}_{310}$\\$\text{TE}_{211}$\\$\text{TE}_{011}$}&
\thead{11.588\\10.267\\10.783}&
\thead{$121$\\$795$\\$4.7\times10^3$}&
\thead{$108$\\$600$\\$863$}&
\thead{$3.0\times10^{-3}$\\$2.5\times10^{-3}$\\$1.8\times10^{-3}$}&
\thead{$4.3\times10^{-8}$\\$1.6\times10^{-8}$\\$6.7\times10^{-10}$}&
\thead{$0.10$\\$5.2\times10^{-5}$\\$1.6\times10^{-5}$}&
\thead{$0.63\pm0.028$\\$\leq0.015$\\$0.070\pm0.005$}\\
\hline
E4(d)& \thead{6061Al\\DT25}&
\thead{$\text{TM}_{010}$\\$\text{TE}_{111}$\\$\text{TE}_{011}$}&
\thead{4.839\\8.482\\10.759}&
\thead{$4.4\times10^{-2}$\\$17$\\$51$}&
\thead{$0.25$\\$32$\\$55$}&
\thead{$6.1\times10^{-3}$\\$2.8\times10^{-3}$\\$1.8\times10^{-3}$}&
\thead{$2.8\times10^{-8}$\\$0.8\times10^{-8}$\\$6.7\times10^{-10}$}&
\thead{$0.15$\\$1.5\times10^{-5}$\\$1.6\times10^{-5}$}&
\thead{$9.65\pm0.56$\\$\leq0.73$\\$26.9\pm1.4$}\\
\hline
E4(eb)& \thead{6061Al\\DT25\\ebAl}&
\thead{$\text{TM}_{310}$\\$\text{TE}_{311}$\\$\text{TE}_{011}$}&
\thead{11.573\\12.002\\10.759}&
\thead{$3.8\times10^3$\\$3.6\times10^4$\\$1.4\times10^4$}&
\thead{$87$\\$302$\\$249$}&
\thead{$3.0\times10^{-3}$\\$2.4\times10^{-3}$\\$1.8\times10^{-3}$}&
\thead{$4.3\times10^{-8}$\\$2.5\times10^{-8}$\\$6.7\times10^{-10}$}&
\thead{$0.10$\\$7.0\times10^{-5}$\\$1.6\times10^{-5}$}&
\thead{$1.62\pm0.06$\\$\leq0.024$\\$0.065\pm0.006$}\\
\hline
E4(sp)& \thead{6061Al\\DT25\\spAl}&
\thead{$\text{TM}_{210}$\\$\text{TE}_{311}$\\$\text{TE}_{011}$}&
\thead{9.457\\12.002\\10.758}&
\thead{$2.6\times10^3$\\$2.9\times10^3$\\$4.5\times10^3$}&
\thead{$536$\\$443$\\$424$}&
\thead{$3.6\times10^{-3}$\\$2.4\times10^{-3}$\\$1.8\times10^{-3}$}&
\thead{$3.9\times10^{-8}$\\$2.5\times10^{-8}$\\$6.7\times10^{-10}$}&
\thead{$0.11$\\$7.0\times10^{-5}$\\$1.6\times10^{-5}$}&
\thead{$\leq0.52$\\$\leq0.048$\\$\leq0.017$}\\
\end{tabular}
\end{ruledtabular}
\end{table*}
 
\bibliography{reference}

\widetext
\clearpage

\begin{center}
\textbf{\large Supplemental Materials: Microwave loss characterization using multi-mode superconducting resonators}
\end{center}
\setcounter{equation}{0}
\setcounter{figure}{0}
\setcounter{table}{0}
\setcounter{page}{1}
\setcounter{section}{0}
\makeatletter
\renewcommand{\theequation}{S\arabic{equation}}
\renewcommand{\thefigure}{S\arabic{figure}}
\renewcommand{\bibnumfmt}[1]{[S#1]}
\renewcommand{\citenumfont}[1]{S#1}
\renewcommand{\thetable}{S\arabic{table}}
\renewcommand{\thesection}{S-\Roman{section}}

\section{Resonant modes in FWGMR}
\begin{figure}
\includegraphics{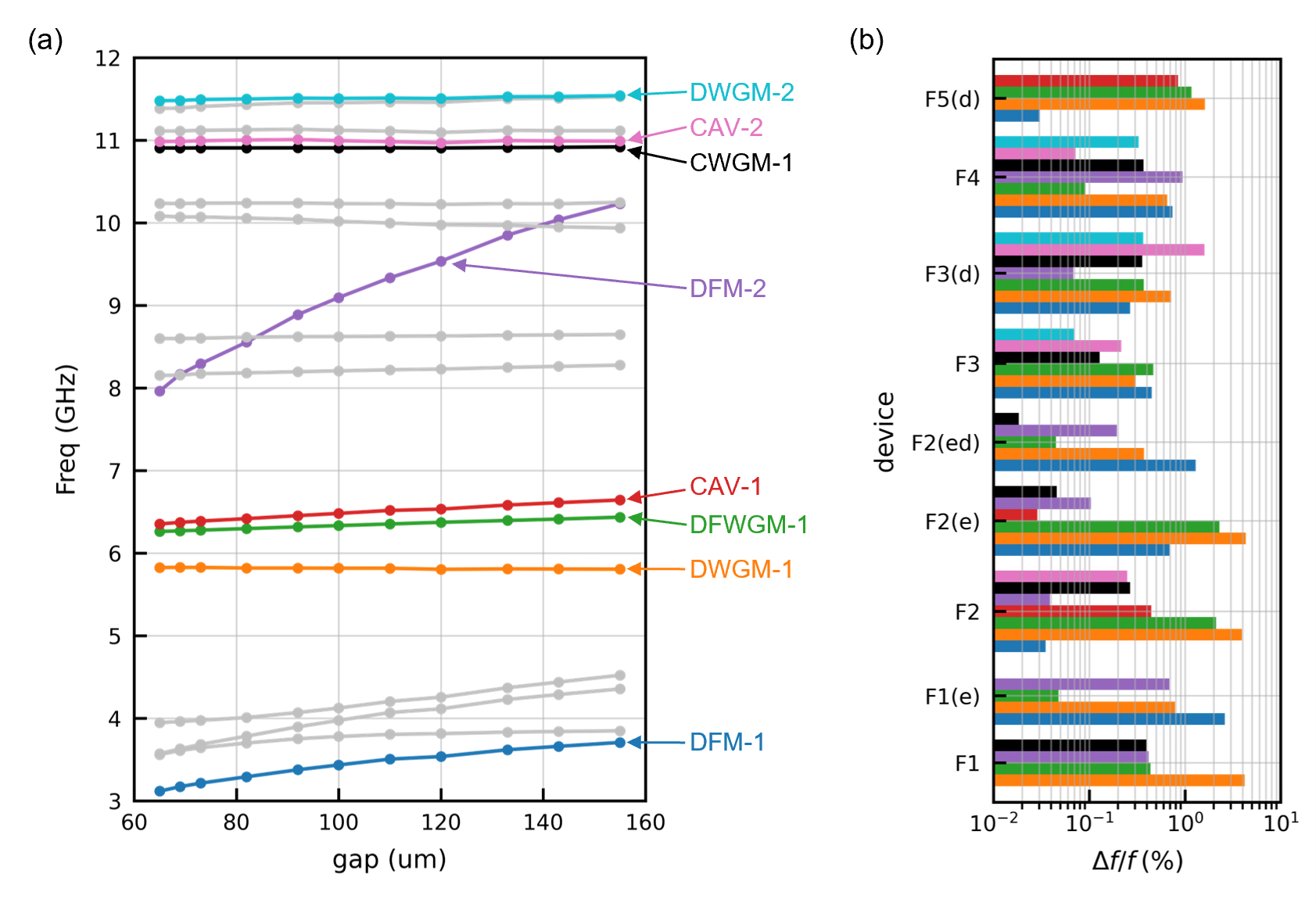}
  \caption{(a) Frequencies of the resonant modes of the FWGMR as a function of the gap between the two planar components. Colored lines and symbols are the modes used in material loss characterization in this work. (b) Fractional difference between the measured and simulated resonance frequencies of all the FWGMRs studied in this work. The color code of the mode is the same as in (a).
  }
  \label{fig:suppl_FWGMR_gap_vs_freq}
\end{figure}

In this work, we calculate the frequencies and the electromagnetic fields of the resonant modes in the FWGMR with finite-element electromagnetic simulation. 
Fig. \ref{fig:suppl_FWGMR_gap_vs_freq}(a) shows the resonance frequencies of the modes as a function of the gap size. The frequencies of the DFMs are very sensitive to the gap size, whereas the frequencies of the DWGMs ,CWGMs, and CAVs are independent of the gap size. As discussed in section \ref{sec:FWGMR} in the main text, we determine the gap size of the devices by matching the measured resonant frequencies with the simulated resonant frequencies, which are matched within $5\%$ for the FWGMRs studied in this work (Fig. \ref{fig:suppl_FWGMR_gap_vs_freq}(b)).

\begin{figure*}
  \includegraphics{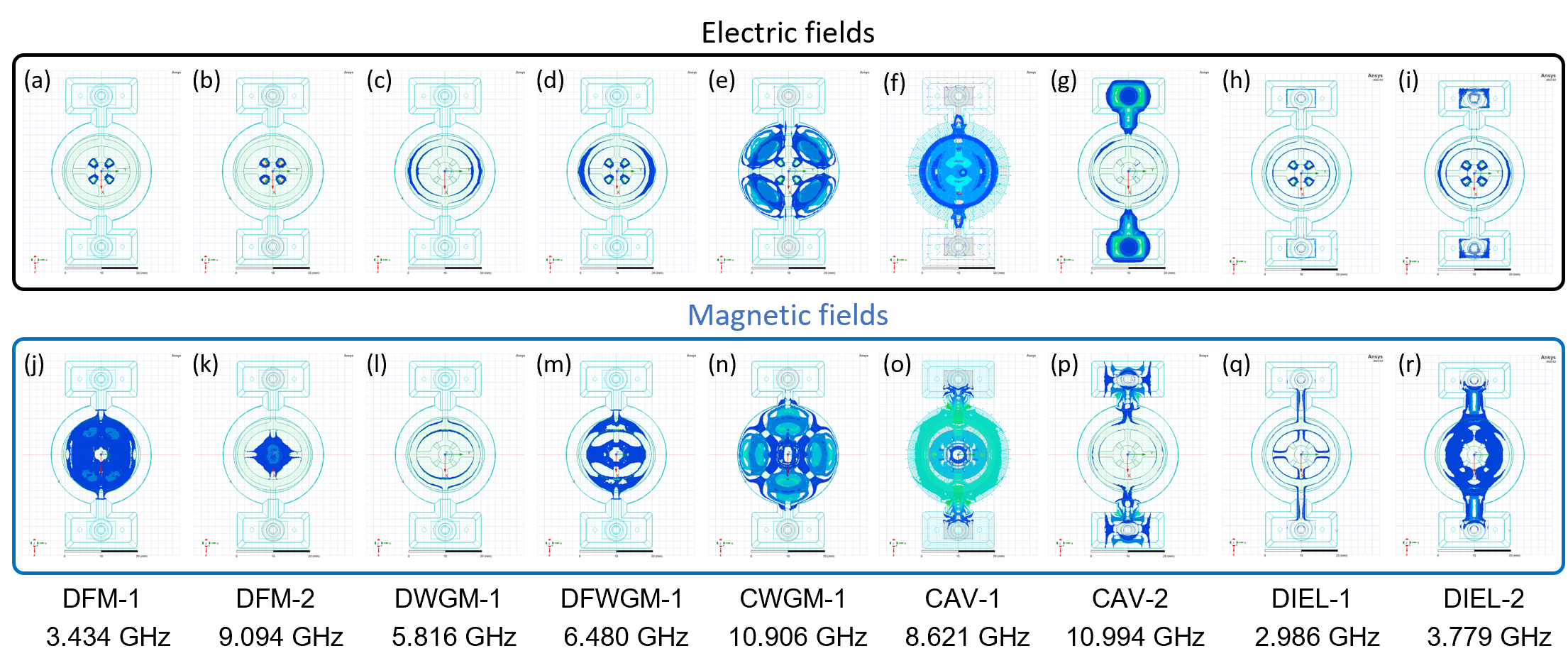}
  \caption{Electric and magnetic fields of selected resonance modes of a FWGMR with 100 $\upmu\text{m}$ gap from finite-element simulation. The isosurfaces in (a-i) represent the magnitude of the electric fields. The isosurfaces in (j-r) represent the magnitude of the magnetic fields.
  }
  \label{fig:suppl_FWGMR_modes}
\end{figure*}

\begin{table*}[b] \centering
\caption{
\label{tab:suppl_FWGMR_particip}
The frequency and the participation factors of selected modes in a FWGMR with $100\upmu\text{m}$ gap. Here, $1/\mathcal{G}$ is the inverse geometric factor of the cavity, $p_{\textrm{MA}}$ is the surface dielectric participation factor of the metal-air (MA) interface, $y_{\textrm{seam}}$ is the seam admittance per unit length of the cavity seam, and $p_{\textrm{diel}}$ is the bulk dielectric participation of the dielectric screws and washers in the assembly.
}
\begin{ruledtabular}
\begin{tabular}{cccccc}
 Mode&
 Freq (GHz)&
$1/\mathcal{G}\,(1/\Omega)$&
 $p_{\textrm{MA}}$&
 $y_{\textrm{seam}}\,(/\Omega/\textrm{m})$&
 $p_{\textrm{diel}}$
 \\
\hline
DFM-1&3.434&$5.1\times10^{-2}$& $2.7\times10^{-6}$&$6.4\times10^{-4}$&
$1.7\times10^{-7}$\\
\hline
DFM-2&9.094&$8.9\times10^{-3}$&$3.5\times10^{-6}$&$7.1\times10^{-5}$&
$2.3\times10^{-8}$\\
\hline
DWGM-1&5.816&$2.8\times10^{-1}$&$3.8\times10^{-6}$&$2.7\times10^{-4}$&
$8.2\times10^{-8}$\\
\hline
DWGM-2&11.506&$1.8\times10^{-1}$&$5.0\times10^{-6}$&$1.7\times10^{-4}$&
$7.9\times10^{-6}$\\
\hline
DFWGM-1&6.480&$6.7\times10^{-2}$&$1.6\times10^{-6}$&$2.0\times10^{-4}$&
$3.2\times10^{-7}$\\
\hline
CWGM-1&10.906&$5.5\times10^{-3}$&$1.5\times10^{-7}$&$2.1\times10^{-3}$&
$5.3\times10^{-5}$\\
\hline
CAV-1&8.621&$6.7\times10^{-3}$&$2.7\times10^{-7}$&$3.2\times10^{-1}$&
$1.5\times10^{-3}$\\
\hline
CAV-2&10.994&$2.1\times10^{-2}$&$8.2\times10^{-7}$&$1.0$&
$3.1\times10^{-1}$\\
\hline
DIEL-1&2.986&$5.2\times10^{-1}$&$3.7\times10^{-6}$&$2.8\times10^{-2}$&
$2.7\times10^{-1}$\\
\hline
DIEL-2&3.779&$1.1\times10^{-1}$&$1.6\times10^{-6}$&$1.2\times10^{-2}$&
$1.1\times10^{-1}$\\
\end{tabular}
\end{ruledtabular}
\end{table*}

\begin{figure*}
\includegraphics{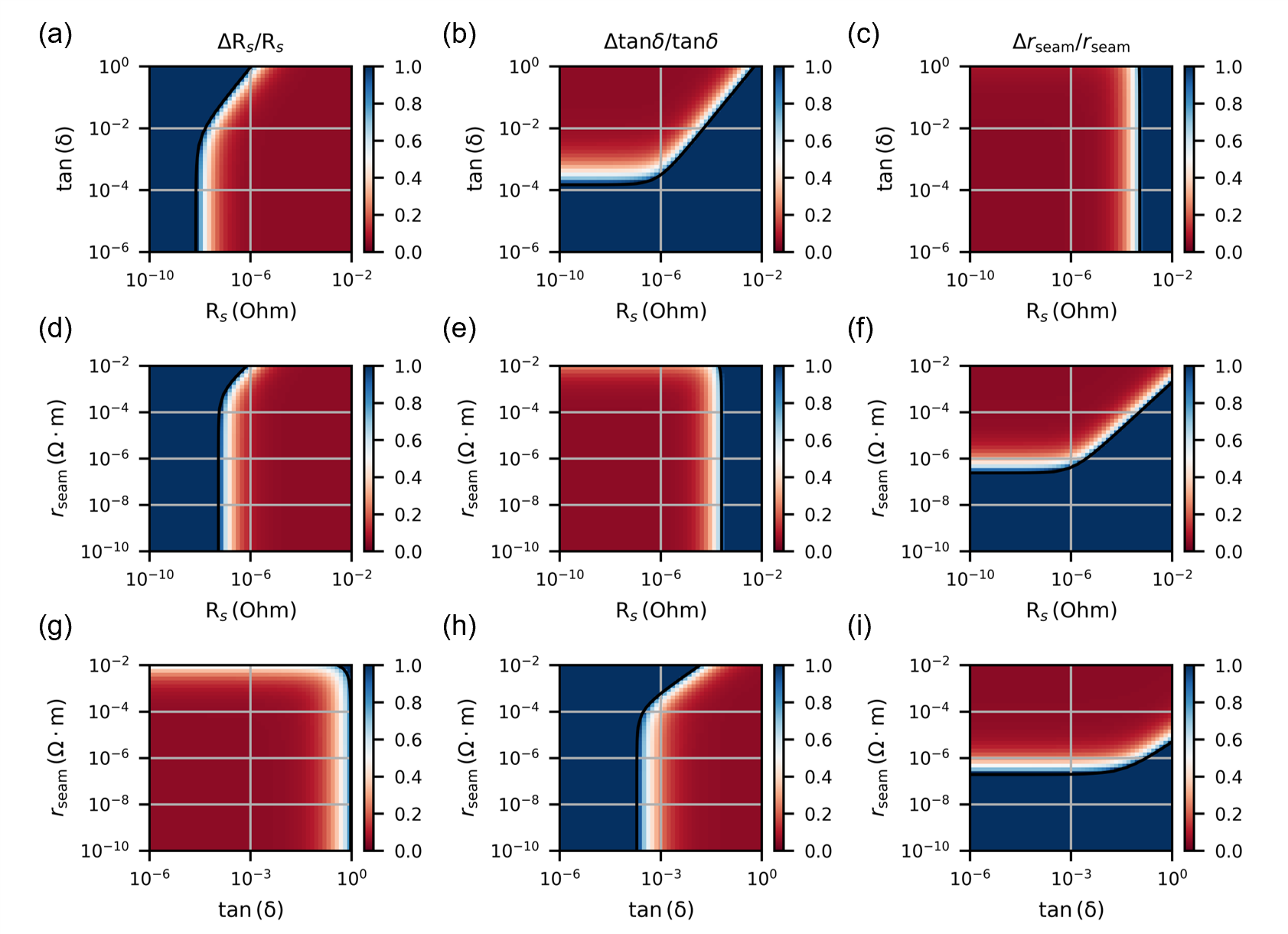}
  \caption{Sensitivity maps of the multi-mode system formed by the DWGM, the DFM, and the CWGM in the FWGMR with 100 $\upmu\text{m}$ gap. (a, b, c) show the system's measurement sensitivity to the material loss factors in the $R_s-\tan{\delta}$ plane evaluated at $r_{\text{seam}} = 10^2\,\upmu\Omega\cdot\text{m}$. (d, e, f) show the system's measurement sensitivity to the material loss factors in the $R_s-r_{\text{seam}}$ plane evaluated at $\tan{\delta} = 5\times10^{-2}$. (g, h, i) show the system's measurement sensitivity to the material loss factors in the $\tan{\delta}-r_{\text{seam}}$ plane evaluated at $R_s = 1\,\upmu\Omega$.
  }
  \label{fig:suppl_FWGMR_5_10_14_senmaps}
\end{figure*}

Fig. \ref{fig:suppl_FWGMR_modes} shows the electromagnetic fields of selected modes in a FWGMR with 100 $\upmu\text{m}$ gap. The loss participation factors of these modes are shown in Table. \ref{tab:suppl_FWGMR_particip}, where $p_{\textrm{diel}}$ is the bulk dielectric participation factor of the dielectric screws and washers used in the assembly, which we approximate as a single dielectric object in the cavity. Fig. \ref{fig:suppl_FWGMR_modes} clearly shows that the differential-whispering-gallery-modes (DWGMs) localized both the electric and the magnetic fields within the gap between the planar components, whereas the differential-fork-modes (DFMs) mostly localize the electric fields within the gap. On the other hand, the electromagnetic fields of the common-whispering-gallery-modes (CWGMs) and the cavity-like modes (CAVs) are more spatially distributed throughout the truncated cylindrical cavity. The very different electromagnetic field distributions make these modes sensitive to different types of microwave losses in the cavity, enabling the multi-mode approach to characterize microwave losses of materials. Fig. \ref{fig:suppl_FWGMR_5_10_14_senmaps} shows all nine projections of sensitivity maps of the multi-mode system formed by the DWGM, the DFM, and the CWGM in a FWGMR with 100 ${\upmu\text{m}}$ gap at the interested range of material loss factors. 

In addition to the modes used in material loss characterization, there are modes that concentrate the electric fields in the dielectric screws and washers (CAV-2, DIEL-1, and DIEL-2). These modes have very large bulk dielectric participation factor ($p_{\textrm{diel}}$ on the order of $10^{-1}$), which can be used to determine the upper bound for the loss tangent of the dielectric components. The lowest upper bound observed in this work is given by the DIEL-2 mode of device F2(e), with internal quality factor equal to $1.2\times10^6$ at $\bar{n} = 6$, which correspond to loss tangent less than $6.9\times10^{-6}$. This upper bound of loss tangent can be used to estimate the bulk dielectric loss contribution from the dielectric components. For most of the modes used in material loss characterization in this work, such as the DWGMs, DFMs, CWGM, and CAV-1, the bulk dielectric loss from the dielectric components contribute to less than $1\%$ of their total internal losses. The only exception is the CAV-2 mode in device F3, the bulk dielectric loss from the dielectric components contributes up to $14\%$ of its total internal loss. 
It's important to note that the loss model used in this work doesn't account for the bulk dielectric loss from the dielectric components, the uncertainty of this additional loss channel is equivalent to extra uncertainty to the measured loss rate. To take this effect into account, the relative uncertainty of the measured loss rate $\epsilon_y$ is set to $20\%$ for the CAV-2 mode in device F3.

\section{Resonant modes in ellipsoidal cavity}
\begin{figure*}
  \includegraphics{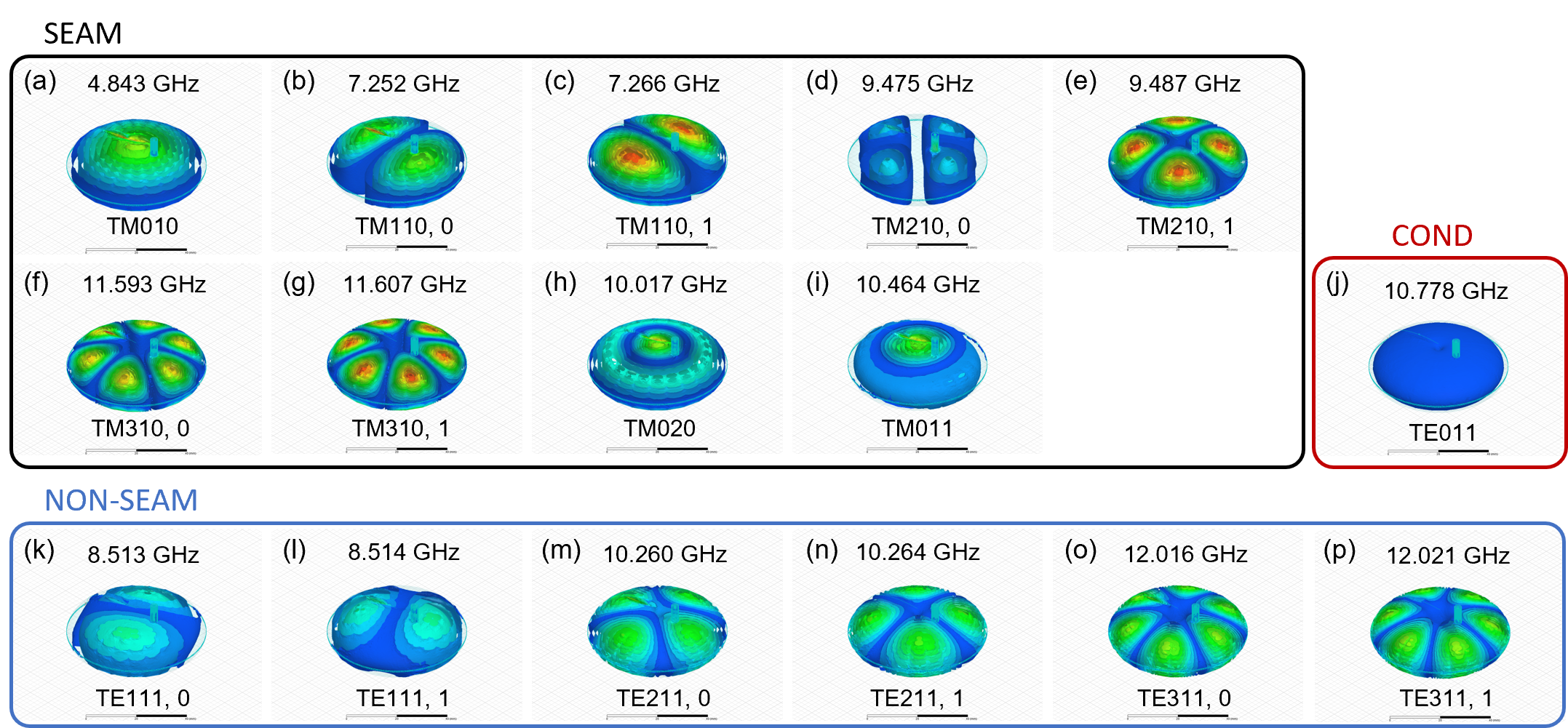}
  \caption{ Electric fields of selected seam-sensitive modes (SEAM), seam-insensitive modes (NON-SEAM), and conductive loss-sensitive mode (COND). The isosurfaces represent the magnitude of the electric fields.
  }
  \label{fig:suppl_ellipscav_modes}
\end{figure*}

\begin{figure*}
  \includegraphics{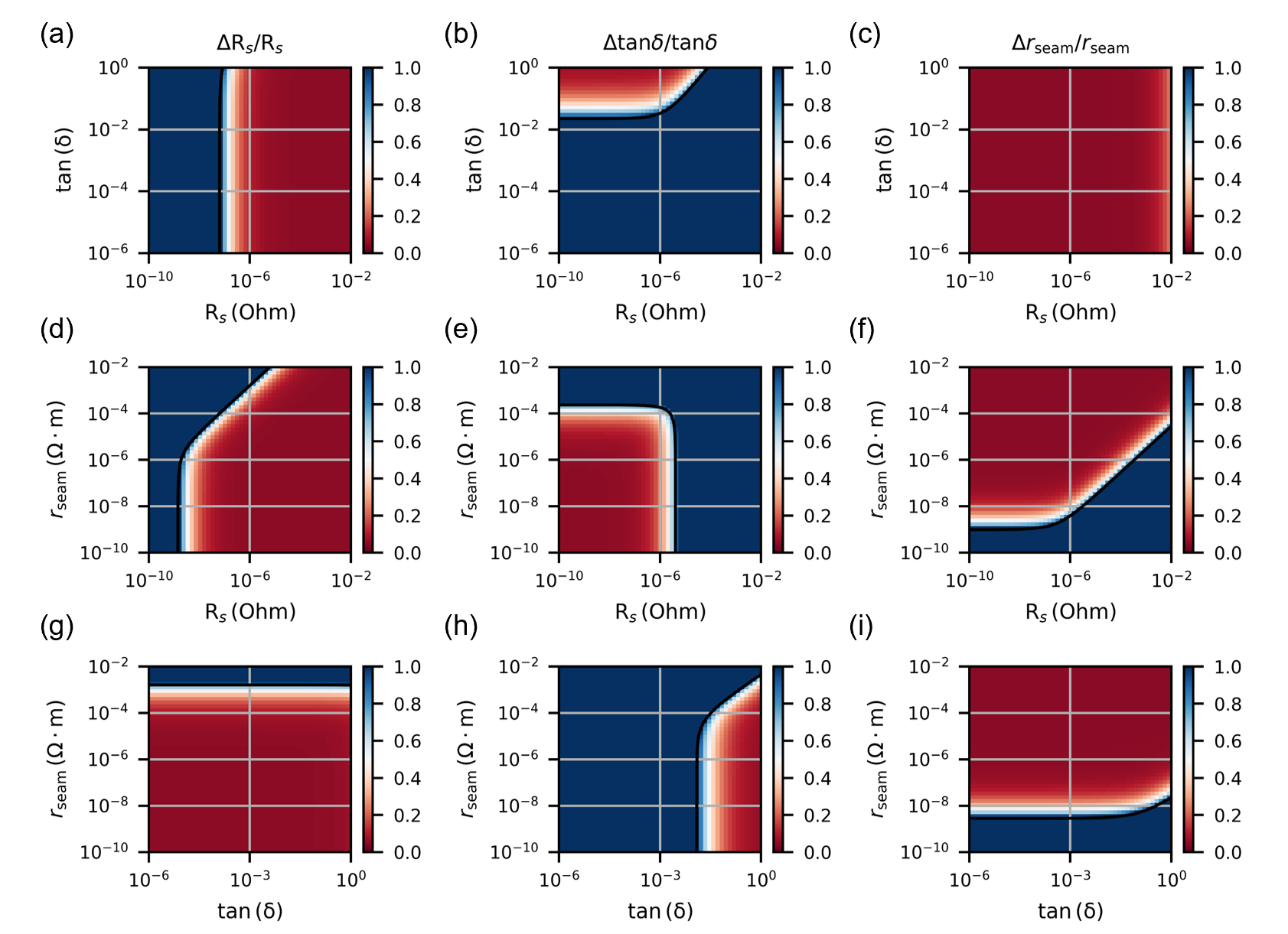}
  \caption{Sensitivity maps of the multi-mode system formed by the seam-sensitive mode, the seam-insensitive mode, and the conductive loss-sensitive mode in the ellipsoidal cavity. (a, b, c) show the system's measurement sensitivity to the material loss factors in the $R_s-\tan{\delta}$ plane evaluated at $r_{\text{seam}} = 10^2\,\upmu\Omega\cdot\text{m}$. (d, e, f) show the system's measurement sensitivity to the material loss factors in the $R_s-r_{\text{seam}}$ plane evaluated at $\tan{\delta} = 5\times10^{-2}$. (g, h, i) show the system's measurement sensitivity to the material loss factors in the $\tan{\delta}-r_{\text{seam}}$ plane evaluated at $R_s = 1\,\upmu\Omega$.
  }
  \label{fig:suppl_ellipscav_3_9_12_senmaps}
\end{figure*}

Similar to the FWGMR, we calculate the frequencies and the electromagnetic fields of the resonant modes in the ellipsoidal cavity with finite-element electromagnetic simulation. The electric fields of the seam loss-sensitive modes, the seam loss-insensitive modes, and the conductive loss-sensitive mode are shown in Fig.~\ref{fig:suppl_ellipscav_modes}. The loss participation factors of selected modes are shown in Table. \ref{tab:suppl_ellipcav_particip}. All nine projections of the sensitivity maps of the multi-mode system formed by the seam-sensitive mode, the seam-insensitive mode, and the conductive loss-sensitive mode are shown in Fig.~\ref{fig:suppl_ellipscav_3_9_12_senmaps}.

\begin{table*}[b] \centering
\caption{
\label{tab:suppl_ellipcav_particip}
Participation factors of selected modes in an ellipsoidal cavity.
}
\begin{ruledtabular}
\begin{tabular}{ccccc}
 Mode&
 Freq (GHz)&
$1/\mathcal{G}\,(1/\Omega)$&
 $p_{\textrm{MA}}$&
 $y_{\textrm{seam}}\,(/\Omega/\textrm{m})$
 \\
\hline
TM010&4.843&$6.1\times10^{-3}$&$2.8\times10^{-8}$&$1.5\times10^{-1}$\\
\hline
TM110&7.252&$4.6\times10^{-3}$&$3.5\times10^{-8}$&$1.3\times10^{-1}$\\
\hline
TM210&9.475&$3.6\times10^{-3}$&$3.9\times10^{-8}$&$1.1\times10^{-1}$\\
\hline
TM020&10.017&$3.3\times10^{-3}$&$3.9\times10^{-8}$&$1.2\times10^{-1}$\\
\hline
TM011&10.464&$2.8\times10^{-3}$&$2.7\times10^{-8}$&$6.3\times10^{-5}$\\
\hline
TE111&8.513&$2.8\times10^{-3}$&$0.8\times10^{-8}$&$1.5\times10^{-5}$\\
\hline
TE211&10.260&$2.5\times10^{-3}$&$1.6\times10^{-8}$&$5.2\times10^{-5}$\\
\hline
TE011&10.778&$1.8\times10^{-3}$&$6.7\times10^{-10}$&$1.6\times10^{-5}$\\
\end{tabular}
\end{ruledtabular}
\end{table*}

\clearpage
\section{Measurement setup}

\begin{figure*}
  \includegraphics{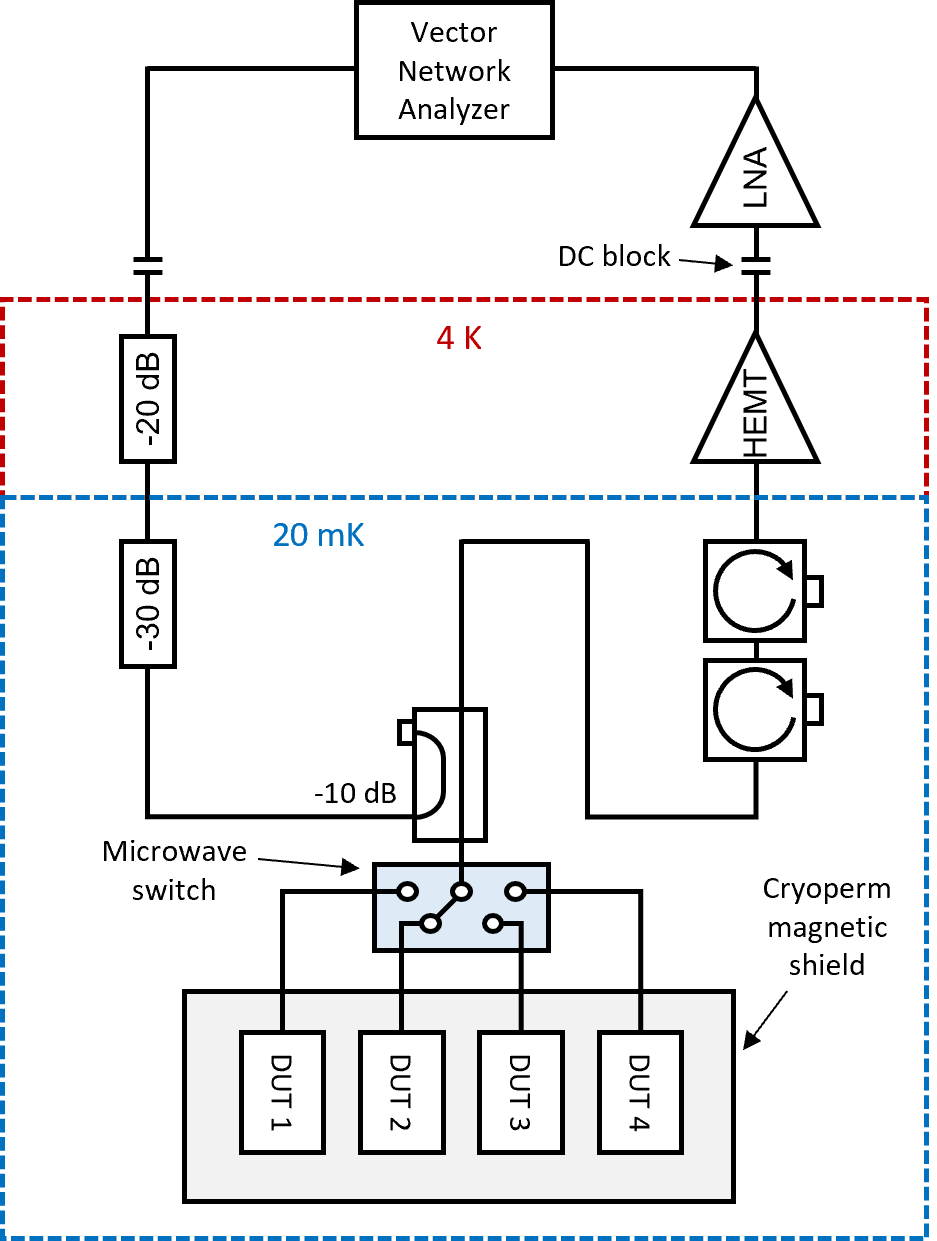}
  \caption{Measurement setup.
  }
  \label{fig:suppl_msmt_setup}
\end{figure*}

Fig.~\ref{fig:suppl_msmt_setup} shows the measurement setup of the resonator measurement. The resonators are installed within a Cryoperm magnetic shield inside a cryogenic-free dilution refrigerator with a base temperature of 20 mK. The input microwave signal passes through a 20 dB and a 30 dB attenuators on the 4K stage and the base stage of the dilution refrigerator, then enters the -10 dB coupling port of a directional coupler and is directed to the coupling port of the sample through a microwave switch. The reflected signal from the sample transmitted through the microwave switch and the transmitted port of the directional coupler, then passes through two isolators at 20 mK and is amplified by a cryogenic HEMT amplifier at 4 K, which is further amplified with a room-temperature low noise amplifier for analysis.

\section{Numerical solution and Monte-Carlo analysis}

\begin{figure*}
  \includegraphics{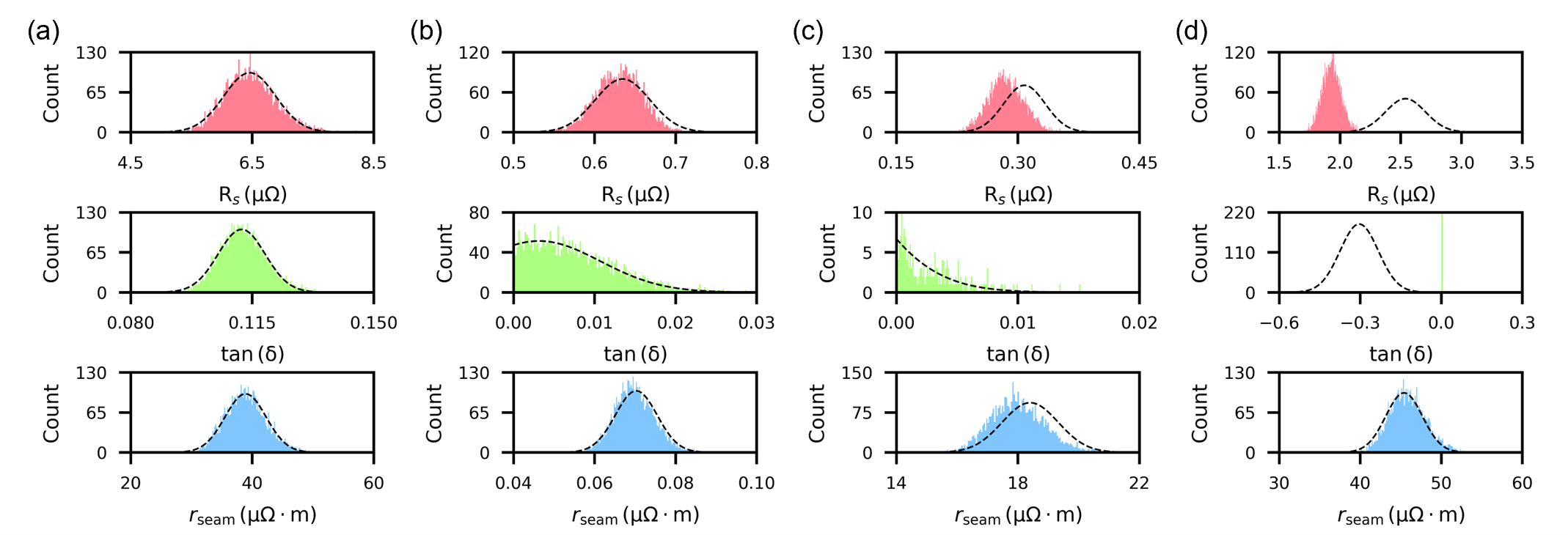}
  \caption{Statistical distributions of the material loss factors for device F4 (a), E3(eb) (b), E1(e) (c), and E1 (d). The histograms are the results of the Monte-Carlo analysis with non-negative least-squares solution. The black dotted curves are the  results of the linear least-squares analytic solutions.
  }
  \label{fig:suppl_material_losses_histogram}
\end{figure*}

As discussed in section \ref{sec:loss model} in the main text, the analytical linear least-squares algorithm does not ensure the extracted material loss factors be non-negative. In order to impose this restriction, we use a numerical non-negative least-squares algorithm and pair it with a Monte-Carlo analysis to extract the loss factors and estimate their variance. 
For each mode, we generate a distribution of the internal quality factors using normal distribution with means and standard deviations equal to the measured values and measurement uncertainties. We take 5000 samples from each of these distributions and use nonlinear least squares to extract the loss factors for each set of samples to obtain the distributions of the material loss factors.
Fig.~\ref{fig:suppl_material_losses_histogram} shows the distributions of the material loss factors for selected devices in this work. The colored histograms are the results of the Monte-Carlo numerical approach and the dotted curves are the predicted distributions from the linear least-square solution, which are normal distributions with means equal to the solution Eq.(\ref{eq:soln}) and variances equal to the diagonal elements of the covariance matrix Eq.(\ref{eq:Cov}). 
When the material loss factors are within or not very far below the system's measurement sensitivity, both approach give consistent results (Fig.~\ref{fig:suppl_material_losses_histogram}a,b). 
On the other hand, when one of the material loss factor is far below the system's measurement sensitivity, the analytic solution Eq.(\ref{eq:soln}) could gives negative solution whereas the numerical approach always guarantee non-negative solution (the distribution of the loss tangent in Fig.~\ref{fig:suppl_material_losses_histogram}c,d). 


\section{Surface roughness of different surface finishing}

\begin{figure*}
  \includegraphics{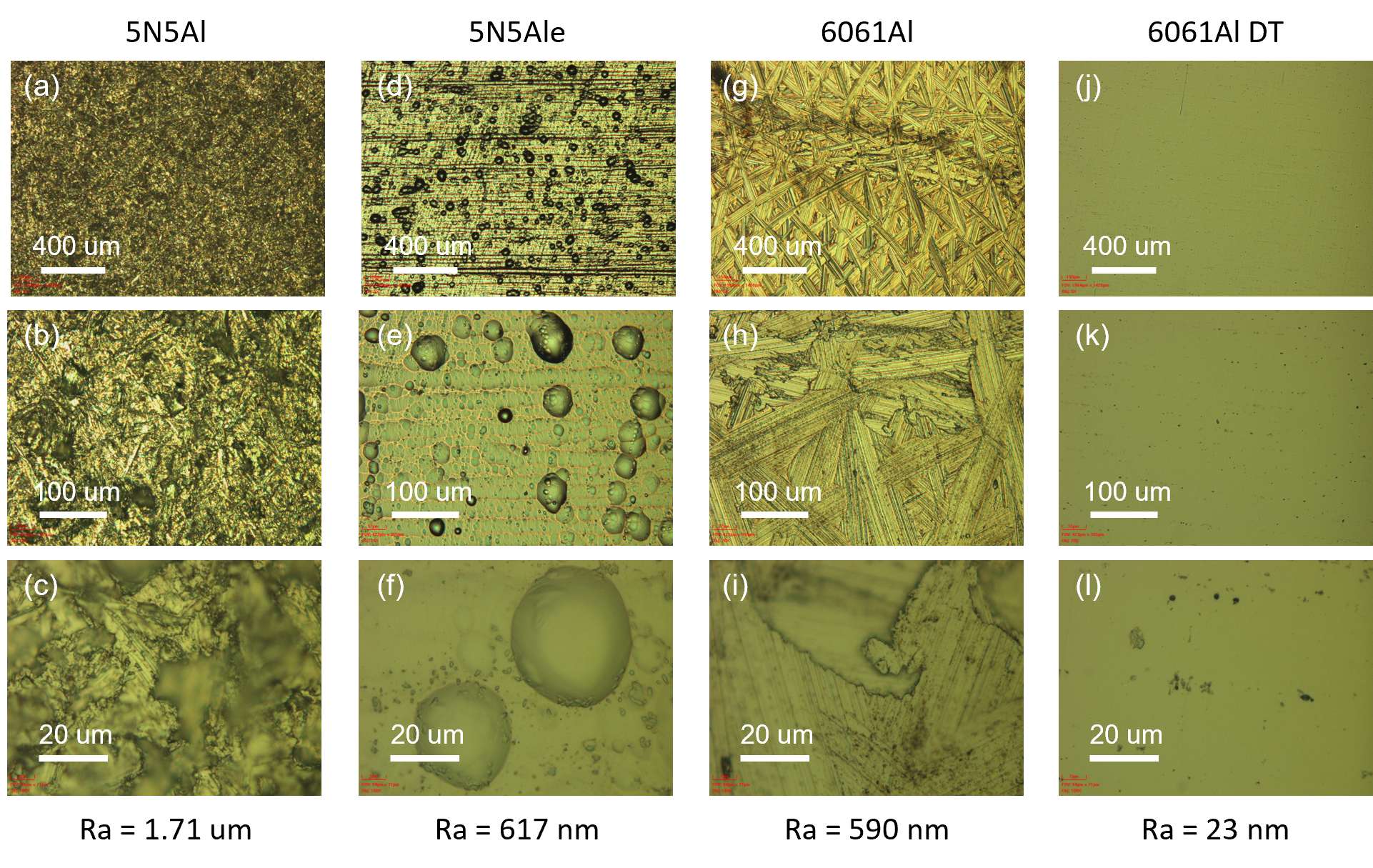}
  \caption{Optical images of the surfaces of as machined 5N5 aluminum (5N5Al) (a, b, c), 5N5 aluminum after chemically etched with Transene aluminum etchatant type A at $50\,^{\circ}\text{C}$ for 2 hours (5N5Ale) (d, e, f), as machined 6061 aluminum alloy (6061Al) (g, h, i), and 6061 aluminum alloy after diamond turning (j, k, l). The average surface roughness (Ra) of these samples are measured by a stylus profilometer with 2 mm scan length.
  }
  \label{fig:suppl_surface_finish}
\end{figure*}

Fig.~\ref{fig:suppl_surface_finish} compares the optical images and the average surface roughness (Ra) of the surfaces of 5N5 aluminum (5N5Al), 5N5 aluminum after chemically etched with Transene aluminum etchant type A at $50\,^{\circ}\text{C}$ for 2 hours (5N5Ale), 6061 aluminum alloy (6061Al), 6061 aluminum alloy after diamond turning (6061Al DT). The average surface roughness (Ra) is measured by a stylus profilometer with a 2 mm scan length. Among these samples, the as-machined 5N5 aluminum (5N5Al) surface has the worst surface finishing. The average roughness of its surface is equal to $1.71\,\upmu\text{m}$. The chemical etching process removed approximately $100\,\upmu\text{m}$ of the machined aluminum surface and improved the average surface roughness to 617 nm. It also generates irregular micro-cavities on the aluminum surface (Fig. \ref{fig:suppl_surface_finish} (e, f)), which are probably due to the non-uniform bubble formation during the chemical reaction in the chemical etching process. On the other hand, the as-machined 6061 aluminum alloy has a better surface finish than the 5N5 aluminum. Its average surface roughness is equal to $590\,\text{nm}$, slightly better than the 5N5 aluminum after chemical etching. Its average surface roughness is further reduced to $23\,\text{nm}$ after removing $25\,\upmu\text{m}$ of the surface material with diamond turning, ending up with a mirror-finished surface.

\end{document}